\pdfoutput=1
\documentclass[journal]{IEEEtran}
\usepackage[T1]{fontenc}
\usepackage{cite}
\usepackage{amsmath,amssymb,amsfonts}
\usepackage{algorithm}
\usepackage{algorithmic}
\usepackage{graphicx}
\usepackage{epsfig}
\usepackage{epstopdf}
\usepackage{textcomp}
\usepackage{xcolor}
\usepackage{subfigure}
\usepackage{booktabs}
\usepackage{multirow}
\usepackage{diagbox}
\usepackage{url}
\usepackage{threeparttable}
\usepackage{array}
\makeatletter
\def\UrlAlphabet{%
      \do\a\do\b\do\c\do\d\do\e\do\f\do\g\do\h\do\i\do\j%
      \do\k\do\l\do\m\do\n\do\o\do\p\do\q\do\r\do\s\do\t%
      \do\u\do\v\do\w\do\x\do\y\do\z\do\A\do\B\do\C\do\D%
      \do\E\do\F\do\G\do\H\do\I\do\J\do\K\do\L\do\M\do\N%
      \do\O\do\P\do\Q\do\R\do\S\do\T\do\U\do\V\do\W\do\X%
      \do\Y\do\Z}
\def\UrlDigits{\do\1\do\2\do\3\do\4\do\5\do\6\do\7\do\8\do\9\do\0}
\g@addto@macro{\UrlBreaks}{\UrlOrds}
\g@addto@macro{\UrlBreaks}{\UrlAlphabet}
\g@addto@macro{\UrlBreaks}{\UrlDigits}
\makeatother

\graphicspath{{figures/}}

\begin{document}
\setlength{\textfloatsep}{10pt}
\ifdefined \GramaCheck
  \newcommand{\CheckRmv}[1]{}
  \newcommand{\figref}[1]{Figure 1}%
  \newcommand{\tabref}[1]{Table 1}%
  \newcommand{\secref}[1]{Section 1}
  \newcommand{\algref}[1]{Algorithm 1}
  \renewcommand{\eqref}[1]{Equation 1}
\else
  \newcommand{\CheckRmv}[1]{#1}
  \newcommand{\figref}[1]{Fig.~\ref{#1}}%
  \newcommand{\tabref}[1]{Table~\ref{#1}}%
  \newcommand{\secref}[1]{Section~\ref{#1}}
  \newcommand{\algref}[1]{Algorithm~\ref{#1}}
  \renewcommand{\eqref}[1]{(\ref{#1})}
\fi
\newtheorem{theorem}{Theorem}
\newtheorem{proposition}{Proposition}
\newtheorem{assumption}{Assumption}
\newtheorem{definition}{Definition}
\newtheorem{condition}{Condition}
\newtheorem{property}{Property}
\newtheorem{remark}{Remark}
\newtheorem{lemma}{Lemma}
\newtheorem{corollary}{Corollary}

\title{Learning-Aided Iterative Receiver for Superimposed Pilots: Design and Experimental Evaluation}

\author{Xinjie~Li,
        Xingyu~Zhou,
        Yixiao~Cao, \IEEEmembership{Graduate Student Member,~IEEE,}
        Jing~Zhang, \IEEEmembership{Member,~IEEE,}
        Chao-Kai~Wen, \IEEEmembership{Fellow,~IEEE,}
        Xiao~Li, \IEEEmembership{Member,~IEEE,}
        and Shi~Jin, \IEEEmembership{Fellow,~IEEE}
\thanks{X. Li, X. Zhou, Y. Cao, J. Zhang, X. Li, and S. Jin are with the National Mobile Communications Research Laboratory, Southeast University, Nanjing 210096, China
(e-mail: lixinjie@seu.edu.cn; \protect \url{xy_zhou@seu.edu.cn}; 220240883@seu.edu.cn; jingzhang@seu.edu.cn; \protect \url{li_xiao@seu.edu.cn}; jinshi@seu.edu.cn).}
\thanks{C.-K. Wen is with Institute of Communications Engineering, National Sun Yat-sen University, Kaohsiung 80424, Taiwan
(e-mail: chaokai.wen@mail.nsysu.edu.tw).}
\thanks{This paper was presented in part at the 2025 IEEE Wireless Communications and Networking Conference (WCNC) \cite{DLDenoiseLi2}.}
}

\maketitle

\begin{abstract}
The superimposed pilot transmission scheme offers substantial potential for improving spectral efficiency in MIMO-OFDM systems, but it presents significant challenges for receiver design due to pilot contamination and data interference. To address these issues, we propose an advanced iterative receiver based on joint channel estimation, detection, and decoding, which refines the receiver outputs through iterative feedback. The proposed receiver incorporates two adaptive channel estimation strategies to enhance robustness under time-varying and mismatched channel conditions. First, a variational message passing (VMP) method and its low-complexity variant (VMP-L) are introduced to perform inference without relying on time-domain correlation. Second, a deep learning (DL) based estimator is developed, featuring a convolutional neural network with a despreading module and an attention mechanism to extract and fuse relevant channel features. Extensive simulations under multi-stream and high-mobility scenarios demonstrate that the proposed receiver consistently outperforms conventional orthogonal pilot baselines in both throughput and block error rate. Moreover, over-the-air experiments validate the practical effectiveness of the proposed design. Among the methods, the DL based estimator achieves a favorable trade-off between performance and complexity, highlighting its suitability for real-world deployment in dynamic wireless environments.
\end{abstract}


\IEEEpeerreviewmaketitle

\section{Introduction}
\IEEEPARstart{C}{hannel} state information (CSI) acquisition at the base station (BS) is essential for the performance of multiple-input multiple-output (MIMO) systems. Traditional approaches allocate separate pilot and data symbols within each coherence block to mitigate mutual interference. Orthogonal pilot (OP) sequences are assigned to different users to eliminate inter-user interference \cite{ceLMMSECol, OPMar1, OPMar2}. However, the overhead from dedicated pilot resource elements (REs) reduces spectral efficiency. To overcome this limitation, the superimposed pilot (SIP) transmission scheme has been proposed as a promising alternative. In this scheme, pilot signals are superimposed onto data symbols in the power domain, enabling more REs to be allocated for data transmission \cite{SIPXie, SIPZhou}. The increased pilot density also supports tasks such as CSI acquisition under high-mobility conditions \cite{SIPHuang, SIPShe} and target sensing \cite{SIPXia, SIPGup}. However, the coexistence of pilots and data introduces pilot contamination and data interference. This complexity requires advanced receiver designs to effectively separate the overlapping components.

Recent advances have explored deep learning (DL) techniques for designing SIP receivers \cite{DDAit, DDZou, DDRez, DDXiao}. For example, a convolutional neural network (CNN) based receiver was proposed in \cite{DDAit} to jointly perform channel estimation and demapping by directly inferring information bits from the received signal. This end-to-end approach incorporates learnable constellation and power allocation, enabling optimization without explicit domain knowledge. However, its black-box nature limits scalability to multi-layer MIMO systems, as it cannot effectively model signal superposition across antennas and layers. To address this, subsequent studies \cite{DDZou, DDRez, DDXiao} have investigated DL-aided receiver designs for multi-stream transmissions. In \cite{DDZou}, self-attention mechanisms enhance feature extraction and mitigate inter-layer interference. In \cite{DDRez}, CNN modules refine least squares (LS) based channel estimates and linear minimum mean square error (LMMSE) based detection to suppress cross interference. Similarly, \cite{DDXiao} employs a modular design with an interference cancellation (IC) mechanism to improve demapping through iterative refinement. Nonetheless, the decoupling of pilot and data in these models introduces considerable complexity and training overhead \cite{DDJu}, while their adaptability to diverse scenarios remains largely unverified. These limitations motivate the integration of domain knowledge into DL model design, fostering interest in joint channel estimation, signal detection, and decoding (JCDD) frameworks.

The JCDD framework has been widely studied in receiver design \cite{VMPKir, JCDDSun, JCDDKar}. In the context of SIP-based receivers, the JCDD framework enhances pilot quality by incorporating iterative feedback from data estimates, thereby improving channel estimation and overall system performance during each iteration \cite{SJCDDMa, SJCDDVer, SJCDDJing, SJCDDGan1, SJCDDGan2}. Integrating channel decoding not only facilitates error correction but also supplies extrinsic information that reinforces the feedback loop and mitigates error propagation. However, most existing JCDD implementations rely on block-fading assumptions and Gaussian models for both channel coefficients and transmitted signals. These idealized assumptions enable tractable formulations for optimizing pilot power allocation \cite{SJCDDMa, SJCDDJing} and sensing strategies \cite{SJCDDGan1, SJCDDGan2}. While insightful, such methods often fail to maintain robustness in practical fast-fading scenarios where channel conditions vary rapidly. 

In \cite{SJCDDQian}, the IC concept was incorporated into the JCDD framework for multiple-input multiple-output orthogonal frequency division multiplexing (MIMO-OFDM) receivers using the SIP scheme. Specifically, LS estimation and code-domain multiplexing despreading \cite{DesCite} were used for channel estimation. Rather than solely improving pilot estimation, this method directly subtracts data interference from the received signal using estimated channel and data information. Although effective in fast-fading environments, the performance of the IC-based JCDD receiver is limited by the inaccuracies of LS-based channel estimates. Alternatively, LMMSE estimation can improve interpolation accuracy; however, it requires knowledge of second-order channel statistics, which is challenging to obtain in dynamic wireless environments. 

Inspired by previous studies and current challenges, this paper proposes a novel IC-based iterative JCDD receiver designed for SIP transmission. The proposed receiver aims to mitigate pilot contamination and data interference. It initially employs LMMSE-based channel estimation. To address performance limitations under dynamic channel conditions, we further explore data-aided approaches based on variational message passing (VMP) and DL.
The main contributions of this paper are as follows:
\begin{itemize}
\item An IC-based JCDD receiver architecture is developed to enhance performance in practical wireless environments. The receiver begins with LMMSE-based channel estimation to improve CSI acquisition. It also refines the computation of extrinsic information in the decoding module to reduce error propagation across iterations. This design improves iterative performance and offers a foundation for better generalization and lower computational complexity. 

\item To overcome the sensitivity of LMMSE-based data-aided estimation to mismatched second-order channel statistics, we derive a VMP-based approach using a factorized probabilistic model. VMP reconstructs an equivalent MIMO system without relying on time-correlation assumptions, thereby improving generalization across diverse channel conditions. To reduce the computational cost of high-dimensional variational inference, we propose a low-complexity variant, VMP-L, which achieves significant complexity reduction by decoupling variable nodes during the message-passing process.

\item A DL-based solution is also developed to further enhance generalization and inference efficiency. This design employs a residual CNN with attention mechanisms and a despreading operation to extract and integrate channel features in the time-frequency domain. The lightweight architecture provides high-resolution channel estimates and demonstrates strong adaptability in time-varying environments.

\item Simulation results under multi-stream and high-mobility scenarios show that the proposed SIP receiver significantly outperforms conventional OP receivers. The VMP, VMP-L, and DL-based methods effectively address performance degradation caused by mismatched and time-varying conditions in LMMSE-based estimation. Notably, the DL-based receiver achieves a favorable balance between performance and computational complexity. Furthermore, an over-the-air (OTA) prototype platform is implemented to validate the practical feasibility and effectiveness of the proposed receiver.
\end{itemize}

The remainder of this paper is organized as follows. \secref{sec_system_jcdd} presents the MIMO-OFDM system model and elaborates on the proposed modifications to the IC-based JCDD receiver. \secref{sec_Adaptive} describes the adaptive enhancement techniques, including both VMP-based estimators and a DL-based method. \secref{sec_simulation} provides detailed simulation results, while \secref{sec_ota} discusses the OTA experimental validation. Finally, \secref{sec_conclusion} concludes the paper.

\emph{Notations:} 
Superscripts ${( \cdot )^{\mathrm{T}}}$ and ${( \cdot )^{\mathrm{H}}}$ denote the transpose and conjugate transpose, respectively. The expectation operator is denoted by $\mathbb{E}\{ \cdot \}$. $\mathbf{I}$ is the identity matrix, $\mathbf{0}$ and $\mathbf{1}$ represent the all-zero matrix and the all-one matrix, respectively. $\text{Diag} ( \mathbf{x} )$ returns a diagonal matrix with $\mathbf{x}$ on the main diagonal, and $\text{diag} ( \mathbf{X} )$ denotes the vectorized diagonal of matrix $\mathbf{X}$. $\mathbf{A}=[a_{i,j}]$ represents a matrix $\mathbf{A}$, where the $(i,j)$-th element is $a_{i,j}$. In addition, $\otimes$ represents the Kronecker product.

\section{System Model and Iterative JCDD Receiver}
\label{sec_system_jcdd}

\begin{figure*}[!t]
  \centering
  \includegraphics[width=0.72\textwidth]{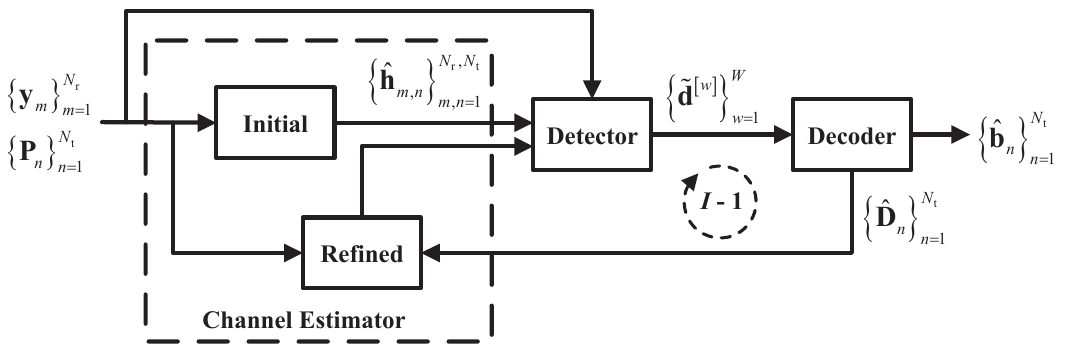}
  \caption{Illustration of the iterative JCDD receiver with $I$ iterations.}
  \label{fig_jcdd}
  \vspace{-1em}
\end{figure*}

\subsection{System Model}
\label{sec_system_model}
Consider a MIMO system with $N_\text{r}$ receive antennas and $N_\text{t}$ transmit antennas. The OFDM frame, consisting of $K$ subcarriers and $T$ consecutive symbols, is transmitted from each antenna. For CSI acquisition, a length-$W$ pilot sequence is mapped onto the time-frequency REs, where $W = KT$. The $W$-point DFT matrix is used to generate the pilot vector ${{\mathbf{p}}_{n}}$ for the $n$-th transmit antenna.

For data transmission, the channel encoder converts $N_\text{b}$ data bits $\mathbf{b}_n \in \{ 0, 1 \}^{N_\text{b}}$ into coded bits with a code rate $R = N_\text{b} / N_\text{c}$ at the $n$-th transmit antenna. The codeword $\mathbf{c}_n \in \{ 0, 1 \}^{N_\text{c}}$ is modulated into a data vector ${{\mathbf{d}}_{n}} \in \mathcal{A}^{W}$ using a complex $M$-ary quadrature amplitude modulation (QAM) constellation $\mathcal{A}$, where $Q = \log_2 M$ is the number of bits per complex symbol.

In the SIP transmission scheme, the transmitted block from the $n$-th antenna is represented as
\begin{equation}
       {{\mathbf{s}}_n} = \sqrt \rho \, {{\mathbf{p}}_n} + \sqrt {1 - \rho } \, {{\mathbf{d}}_n},
       \label{equ_s_n}
\end{equation} 
where $\rho$ is the power allocation factor. After the OFDM receiving procedure is performed, signals from $W$ time-frequency REs are collected. Specifically, the received signal at the $m$-th antenna, ${{\mathbf{y}}_m} \in {\mathbb{C}^{W \times 1}}$, can be expressed as
\begin{equation}
       \label{equ_y_m}
       {{\mathbf{y}}_m} = \sum\limits_{n = 1}^{{N_{\text{t}}}} { {\left( {\sqrt{\rho} \, {{\mathbf{P}}_n} + \sqrt{1 - \rho } \, {{\mathbf{D}}_n}} \right)} {{\mathbf{h}}_{m,n}} }  + {{\mathbf{w}}_m},
\end{equation}
where ${{\mathbf{P}}_n} = {\text{Diag}}( {{{\mathbf{p}}_n}} )$, ${{\mathbf{D}}_n} = {\text{Diag}}( {{{\mathbf{d}}_n}} )$, ${{\mathbf{h}}_{m,n}}$ represents the channel vector corresponding to the $m$-th receive antenna and the $n$-th transmit antenna, and ${{\mathbf{w}}_m}$ denotes the additive white Gaussian noise (AWGN) with variance $\sigma _w^2$.

Additionally, ${{\mathbf{y}}_m}$  can be indexed separately in the time and frequency domains, and the corresponding received components are reshaped into a matrix ${{\mathbf{Y}}_m} \in {\mathbb{C}^{K \times T}}$. The indexing applies to the channel vector ${{\mathbf{h}}_{m,n}}$ and transmitted symbol vector ${{\mathbf{s}}_{n}}$, denoted by ${{\mathbf{H}}_{m,n}}$ and ${{\mathbf{S}}_{n}}$, respectively. The $t$-th column ($t \in \{1, \ldots, T\}$) of ${{\mathbf{Y}}_m}$, denoted as ${{\mathbf{y}}_m^{(t)}} \in \mathbb{C}^{K \times 1}$, can be represented as
\begin{equation}
       \label{equ_y_m_t}
       {\mathbf{y}}_m^{\left( t \right)} = \sum\limits_{n = 1}^{{N_{\text{t}}}} {{\text{Diag}}\left( {{\mathbf{s}}_n^{\left( t \right)}} \right) \cdot {\mathbf{h}}_{m,n}^{\left( t \right)}}  + {\mathbf{w}}_m^{\left( t \right)},
\end{equation}
where ${{\mathbf{h}}_{m,n}^{( t )}}$ refers to the $t$-th column of ${{\mathbf{H}}_{m,n}}$, and ${{\mathbf{s}}_n^{( t )}}$ refers to the $t$-th column of ${{\mathbf{S}}_{n}}$.

At the $w$-th RE, the received components of ${{\mathbf{y}}_m}$ ($m=1,\ldots,N_{\text{r}}$) can be collected as a length-$N_\text{r}$ vector, denoted by ${{\mathbf{y}}^{[ w ]}} = {[ {{y_m^w}} ]} \in \mathbb{C}^{{N_{\text{r}}} \times 1}$. An equivalent MIMO subsystem can be accordingly expressed as
\begin{equation}
       \label{equ_yw_mimo}
       {{\mathbf{y}}^{\left[ w \right]}} = {{\mathbf{H}}^{\left[ w \right]}}\left( {\sqrt \rho \, {{\mathbf{p}}^{\left[ w \right]}} + \sqrt {1 - \rho } \, {{\mathbf{d}}^{\left[ w \right]}}} \right) + {{\mathbf{w}}^{\left[ w \right]}},
\end{equation}
where ${{\mathbf{H}}^{[ w ]}} = [ h_{m,n}^w ] \in \mathbb{C}^{{N_{\text{r}}} \times {N_{\text{t}}}}$ is the channel matrix, ${{\mathbf{p}}^{[ w ]}} = {[ p_n^w ] \in \mathbb{C}^{{N_{\text{t}}} \times 1}}$ is the pilot vector, ${{\mathbf{d}}^{[ w ]}} = {[ d_n^w ] \in \mathbb{C}^{{N_{\text{t}}} \times 1}}$ is the data vector, and ${{\mathbf{w}}^{[ w ]}}$ is the additive noise.

\vspace{-1em}
\subsection{Modified IC-based JCDD Receiver} 
In this subsection, we present a modification of the IC-based iterative receiver proposed in \cite{SJCDDQian}. Specifically, we enhance the channel estimation process by employing LMMSE-based estimation to improve accuracy, and we refine the calculation of log-likelihood ratios (LLRs) for the decoder. As shown in \figref{fig_jcdd}, the modified JCDD design includes three main components: channel estimation, signal detection, and decoding. These components are executed iteratively over $I$ iterations. 

\subsubsection{Channel Estimation}
\label{sec_Iterative_CE}
At the first JCDD iteration ($i=1$), initial channel estimates are obtained using the LMMSE-based method: 
\begin{subequations}
       \label{equ_CE_lmmse_initial}
       \begin{align}
          {\mathbf{h}}_{m,n}^{{\text{LS, init}}} &= {{{\left( {\sqrt \rho \, {{\mathbf{P}}_n}} \right)}^{ - 1}}{{\mathbf{y}}_m}}, \\
          {\mathbf{h}}_{m,n}^{{\text{LMMSE}}} &= {{\mathbf{A}}_{{\text{LMMSE}}}}{\mathbf{h}}_{m,n}^{{\text{LS, init}}},
       \end{align}
\end{subequations}
where ${{\mathbf{A}}_{{\text{LMMSE}}}}={{\mathbf{A}}_{{\text{LMMSE}}}^{{\text{Time}}}}\otimes{{\mathbf{A}}_{{\text{LMMSE}}}^{{\text{Freq}}}}$ denotes the LMMSE-based interpolation in the time and frequency domains. These interpolation matrices are defined as
\begin{subequations}
       \label{equ_Almmse}
       \begin{align}
         \label{equ_Almmse_time}
          {{\mathbf{A}}_{{\text{LMMSE}}}^{{\text{Time}}}} &= {{\mathbf{R}}_{{\mathbf{hh}}}^{{\text{Time}}}}{ {\left( {{{\mathbf{R}}_{{\mathbf{hh}}}^{{\text{Time}}}} + \sigma _w^2{\mathbf{I}}} \right)}^{ - 1}}, \\
          {{\mathbf{A}}_{{\text{LMMSE}}}^{{\text{Freq}}}} &= {{\mathbf{R}}_{{\mathbf{hh}}}^{{\text{Freq}}}}{ {\left( {{{\mathbf{R}}_{{\mathbf{hh}}}^{{\text{Freq}}}} + \sigma _w^2{\mathbf{I}}} \right)}^{ - 1}},
       \end{align}
\end{subequations}
where ${{\mathbf{R}}_{{\mathbf{hh}}}^{{\text{Time}}}}$ and ${{\mathbf{R}}_{{\mathbf{hh}}}^{{\text{Freq}}}}$ are the channel correlation matrices in the time and frequency domains, respectively.

At the $i$-th iteration (${i > 1}$) of the JCDD process, the refined estimation module uses the pilot sequences $\{ {{{{\mathbf{P}}}_n}} \}_{n = 1}^{{N_{\text{t}}}}$ and soft data estimates $\{ {{{\widehat{\mathbf{D}}}_n}} \}_{n = 1}^{{N_{\text{t}}}}$ to mitigate pilot contamination and data interference from the received signal, thus improving the accuracy of channel estimation. The received component corresponding to the $n$-th transmit antenna is computed as
\begin{equation} 
       \label{equ_yp_hat} 
              {\widehat{\mathbf{y}}}_{\text{p},m,n} = {\mathbf{y}}_m - \sum\limits_{\substack{n' \ne n}} 
              { {\sqrt \rho \, {{\mathbf{P}}_{n'}}} {\widehat{\mathbf{h}}}_{m,n'}} 
               - \sum\limits_{n' = 1}^{{N_{\text{t}}}} {{  {\sqrt {1 - \rho } \, {\widehat{\mathbf{D}}}_{n'}}  } {\widehat{\mathbf{h}}}_{m,n'}}, 
\end{equation}
where ${{{\widehat{\mathbf{h}}}_{m,n}}} = { {{\mathbf{h}}_{m,n}^{{\text{LMMSE}}}} }$ is obtained by the LMMSE-based channel estimates at the ($i-1$)-th JCDD iteration. For notational convenience, $\sum_{\substack{n' \ne n}}$ denotes summation over all ${N_{\text{t}}}$ transmit antennas except the $n$-th one.
The refined LMMSE-based estimation is calculated as
\begin{subequations}
       \label{equ_CE_lmmse_refined}
       \begin{align}
         \label{equ_CE_ls_refined}
          {\mathbf{h}}_{m,n}^{{\text{LS}}} &= {{{\left( {\sqrt \rho \, {{\mathbf{P}}_n}} \right)}^{ - 1}}{\widehat{\mathbf{y}}}_{\text{p},m,n}}, \\
          {\mathbf{h}}_{m,n}^{{\text{LMMSE}}} &= {{\mathbf{A}}_{{\text{LMMSE}}}}{\mathbf{h}}_{m,n}^{{\text{LS}}}.
       \end{align}
\end{subequations}
 
\subsubsection{Signal Detection}
\label{sec_Iterative_SD} 
Given the channel estimates at the $i$-th JCDD iteration, represented as ${{\widehat{\mathbf{H}}}^{[w]}} = [ {{{\widehat h}_{m,n}^w}} ] \in \mathbb{C}^{{N_{\text{r}}} \times {N_{\text{t}}}}$, the detection module mitigates the pilot interference from the received signal in \eqref{equ_yw_mimo}, i.e., ${{\widehat{\mathbf{y}}}_{{\text{d}}}^{[w]}} = {{\mathbf{y}}^{[w]}} - {{\widehat{\mathbf{H}}}^{[w]}}( {\sqrt \rho  \, {{\mathbf{p}}^{[w]}}} )$. The LMMSE-based detection for data symbols is computed as follows \cite{MMSEICStu}
\begin{subequations}
       \label{equ_SD_lmmse}
       \begin{align}   
          {{\mathbf{G}}^{\left[ w \right]}} &= {\left( {{{\widetilde{\mathbf{H}}}^{\left[ w \right]}}} \right)^{\mathrm{H}}}{\left( {{{\widetilde{\mathbf{H}}}^{\left[ w \right]}}{{\left( {{{\widetilde{\mathbf{H}}}^{\left[ w \right]}}} \right)}^{\mathrm{H}}} + \sigma _w^2{\mathbf{I}}} \right)^{ - 1}}, \\
          {{\widetilde{\mathbf{d}}}^{[w]}} &= {{\text{Diag}}{({\boldsymbol{\mu}^{[w]}})}^{-1}}{{\mathbf{G}}^{[w]} {\widehat{\mathbf{y}}}_{\text{d}}^{[w]}},
       \end{align}
\end{subequations}
where ${{\widetilde{\mathbf{H}}}^{[w]}} = \sqrt {1 - \rho } \, {{\widehat{\mathbf{H}}}^{[w]}}$, and $\boldsymbol{\mu}^{[w]}={{\text{diag}}( {{{\mathbf{G} ^{[w]}}{{\widetilde{\mathbf{H}}}^{[w]}}}} )}$ is a vector consisting of normalization factors. To further improve the reliability of the detected data, the set of detected symbols $\{ {{{\widetilde{\mathbf{d}}}^{[w]}}} \}_{w = 1}^W$ is passed to the decoding module for error correction.

\subsubsection{Decoding}
\label{sec_Iterative_DEC}
The extrinsic LLRs for the $q$-th bit of the transmitted data symbol corresponding to the $n$-th transmit antenna at the $w$-th RE are given by \cite{MMSEICStu}
\begin{equation}
       \label{equ_LLR_e}
       L_{{\text{E}},n,q}^{w} = {{\eta} _{n}^{w}}\left( {\mathop {\min }\limits_{a \in \mathcal{A}_q^{\left( 0 \right)}} {{\left| {{{\widetilde{d}}_{n}^{w}} - a} \right|}^2} - \mathop {\min }\limits_{a \in \mathcal{A}_q^{\left( 1 \right)}} {{\left| {{{\widetilde{d}}_{n}^{w}} - a} \right|}^2}} \right),
\end{equation}
where ${\mathcal{A}_q^{( 0 )}}$ and ${\mathcal{A}_q^{( 1 )}}$ represent the subsets of the constellation set $\mathcal{A}$ in which the $q$-th bit is 0 and 1, respectively. ${{\widetilde{d}}_{n}^{w}}$ is the $n$-th element of ${{\widetilde{\mathbf{d}}}^{[w]}}$, and the term ${\eta _{n}^{w}}={\mu _{n}^{w}} / ({1-{\mu _{n}^{w}}})$ is calculated using the $n$-th element of $\boldsymbol{\mu}^{[w]}$. The extrinsic information is passed to the channel decoder, and the output LLRs are treated as new \emph{a priori} information ${L}_{\text{A},n,q}^{w}$ for soft remapping \cite{SJCDDQian}. The soft data estimates are computed as
\begin{subequations}
       \label{equ_DEC_dhat}
       \begin{align}
         {\text{Pr}}\left( {a|L_{{\text{A}},n}^{w}} \right) &= \frac{{\exp \left( {{l_i}} \right)}}{{\sum\nolimits_{i' = 1}^M {\exp \left( {{l_{i'}}} \right)} }} , \\
         {l_i} &= \sum\limits_{q = 1}^Q {\ln  \left( {\sigma \left( {\left( {2{a_q} - 1} \right)L_{{\text{A}},n,q}^{w}} \right)} \right)} , \\
         {{\widehat{d}}_{n}^{w}} &= \sum\limits_{a \in \mathcal{A}} {a \cdot {\text{Pr}}\left( {a|{L}_{{\text{A}},n}^{w}} \right)} ,
       \end{align}
\end{subequations}
where $\sigma (x)=1/(1+\exp(-x))$  is the sigmoid function, $\{ {{l_i}} \}_{i = 1}^M$ are the logits, and $a_q$ is the $q$-th bit associated with symbol $a$. During the $i$-th iteration ($i < I$) of the JCDD process, the soft data estimate $\widehat {\mathbf{d}}_n=[{{\widehat{d}}_{n}^{w}}] \in \mathbb{C}^{W \times 1}$ is used for refined channel estimation in the ($i+1$)-th iteration. The corresponding diagonal matrix is given by $\widehat {\mathbf{D}}_n=\text{Diag} (\widehat {\mathbf{d}}_n)$. At the final iteration ($i = I$), the decoder directly outputs the uncoded data bits $\widehat {\mathbf{b}}_n$.

\section{Adaptive Enhanced Approach for Refined Channel Estimation}
\label{sec_Adaptive}
Benefiting from the full-grid utilization of superimposed pilots and the iterative JCDD framework, the proposed receiver enables reliable tracking of channel variations and improved demodulation performance. Moreover, the configuration with $I=2$ provides substantial performance gains while maintaining acceptable overhead, as demonstrated in \secref{sec_simulation}. 

However, the current design relies on an LMMSE-based refined channel estimation method, which shows limited generalization across diverse application scenarios due to the dynamic behavior of MIMO-OFDM channels. This limitation stems from the time-domain interpolation defined in \eqref{equ_Almmse_time}, which depends on the time-domain channel covariance matrix ${{\mathbf{R}}_{{\mathbf{hh}}}^{{\text{Time}}}}$. This matrix is typically estimated by averaging second-order moments from available channel samples. When the time-varying characteristics of the training samples differ from those encountered during testing, the estimation accuracy may deteriorate. To address this issue, we propose the following adaptive schemes to enhance the accuracy of channel estimation in time-varying environments.

\subsection{VMP}
\label{sec_Adaptive_VMP} 
According to \eqref{equ_y_m_t}, ${{\mathbf{y}}_m^{( t )}}$ ($m=1, \ldots, N_{\text{r}}$) can be concatenated as ${{\mathbf{y}}^{( t )}} = { [ {{{( {{\mathbf{y}}_1^{( t )}} )}^{\mathrm{T}}}, \ldots,    {{( {{\mathbf{y}}_{{N_{\text{r}}}}^{( t )}} )}^{\mathrm{T}}}}  ]^{\mathrm{T}}} \in {\mathbb{C}^{{N_{\text{r}}}K \times 1}}$, forming an equivalent MIMO system
\begin{subequations}
       \label{equ_y_t}
       \begin{align}
         & {{\mathbf{y}}^{\left( t \right)}} = {{\mathbf{S}}^{\left( t \right)}}{{\mathbf{h}}^{\left( t \right)}} + {{\mathbf{w}}^{\left( t \right)}}, \\
         & {{\mathbf{S}}^{\left( t \right)}} = {{\mathbf{I}}_{{N_{\text{r}}}}} \otimes \left( {\left( {{{\mathbf{1}}_{1 \times {N_{\text{t}}}}} \otimes {{\mathbf{I}}_K}} \right){\text{Diag}}( {{{\mathbf{s}}^{\left( t \right)}}} )} \right) ,
       \end{align}
\end{subequations}
where ${{\mathbf{h}}^{( t )}} = { [ {{{( {{\mathbf{h}}_{1,1}^{( t )}} )}^{\mathrm{T}}}, \ldots,  {{( {{\mathbf{h}}_{{N_{\text{r}}},{N_{\text{t}}}}^{( t )}} )}^{\mathrm{T}}}} ]^{\mathrm{T}}} \in {\mathbb{C}^{{N_{\text{r}}}{N_{\text{t}}}K \times 1}}$, ${{\mathbf{1}}_{1 \times {N_{\text{t}}}}}$ is a length-${N_{\text{t}}}$ all-ones row vector, and the transmitted symbol vector is concatenated as ${{\mathbf{s}}^{( t )}} = { [ {{{( {{\mathbf{s}}_1^{( t )}} )}^{\mathrm{T}}}, \ldots, {{( {{\mathbf{s}}_{{N_{\text{t}}}}^{( t )}} )}^{\mathrm{T}}}}  ]^{\mathrm{T}}} \in {\mathbb{C}^{{N_{\text{t}}}K \times 1}}$.

\begin{figure}[t]
  \centering
  \begin{minipage}{3.2in}
    \centerline{\includegraphics[width=3.2in]{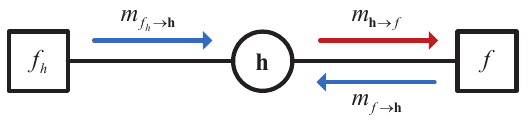}}
  \end{minipage}
  \caption{The channel vector factor graph representation corresponding to the probabilistic model in \eqref{equ_p_ysh}. The variable and factor nodes are represented by circles and rectangles, respectively.}
  \label{fig_vmp_factor}
\end{figure}

For simplicity, omitting the superscript ${(  \cdot  )^{( t )}}$, the joint probability density function (pdf) of the transmitted symbols $\mathbf{s}$, the channel coefficients $\mathbf{h}$, and the observation $\mathbf{y}$ is factorized as
\begin{subequations}
       \label{equ_p_ysh}
       \begin{align}
         & p\left( {{\mathbf{y}},{\mathbf{s}},{\mathbf{h}}} \right) = p\left( {{\mathbf{y}}|{\mathbf{s}},{\mathbf{h}}} \right)p\left( {\mathbf{h}} \right)p\left( {\mathbf{s}} \right), \\
         & f \triangleq p\left( {{\mathbf{y}}|{\mathbf{s}},{\mathbf{h}}} \right) , ~~
         {f_h} \triangleq p\left( {\mathbf{h}} \right).
       \end{align}
\end{subequations}
This probabilistic model can be represented by a factor graph, where the subproblem specific to the channel vector is shown in \figref{fig_vmp_factor}. The channel coefficients to be estimated are denoted by a variable node $\mathbf{h}$, while the factor node $f_h$ represents prior knowledge of the channel model, and the observed factor node $f$ denotes the likelihood probability.

The VMP algorithm approximates the joint pdf in \eqref{equ_p_ysh} with an auxiliary function $b_{\mathbf{h}}$, minimizing the Kullback-Leibler (KL) divergence between $b_{\mathbf{h}}$ and $p( {{\mathbf{y}},{\mathbf{s}},{\mathbf{h}}} )$ \cite{VMPKir}. The estimated auxiliary function, $b_{\mathbf{h}} \propto {m_{{\mathbf{h}} \to f}}$, corresponds to the channel estimates ${\widehat{\mathbf{h}}}$. According to \figref{fig_vmp_factor}, the message from the variable node to the observed factor node ${m_{{\mathbf{h}} \to f}}$ is given by
       \begin{align}
              {m_{{\mathbf{h}} \to f}} &= {m_{f \to {\mathbf{h}}}} \cdot {m_{{f_h} \to {\mathbf{h}}}} \notag \\
              &\propto \exp \left\{ { - {{\left( {{\mathbf{h}} - {\widehat{\mathbf{h}}}} \right)}^{\mathrm{H}}}{{\widehat {\boldsymbol{\Sigma}} }^{ - 1}}\left( {{\mathbf{h}} - {\widehat{\mathbf{h}}}} \right)} \right\}, \label{equ_m_v2f}
       \end{align}
which is proportional to a Gaussian pdf with mean ${\widehat{\mathbf{h}}}$ and covariance ${\widehat {\boldsymbol{\Sigma}} }$. In \eqref{equ_m_v2f}, the prior $f_h$ is calculated as
\begin{equation}
       \label{equ_m_hp2v}
       {m_{{f_h} \to {\mathbf{h}}}} = p\left( {\mathbf{h}} \right) \propto \exp \left\{ { - {{\left( {{\mathbf{h}} - {{\mathbf{h}}_{{\text{p}}}}} \right)}^{\mathrm{H}}}{\boldsymbol{\Sigma}} _{{\text{p}}}^{ - 1}\left( {{\mathbf{h}} - {{\mathbf{h}}_{{\text{p}}}}} \right)} \right\},
\end{equation}
where the prior mean ${{\mathbf{h}}_{{\text{p}}}} = \mathbf{0}$, and the prior covariance ${\boldsymbol{\Sigma}} _{{\text{p}}}={{\mathbf{R}}_{{\mathbf{hh}}}^{{\text{Spat}}}} \otimes {{\mathbf{R}}_{{\mathbf{hh}}}^{{\text{Freq}}}}$, where ${{\mathbf{R}}_{{\mathbf{hh}}}^{{\text{Spat}}}} \in \mathbb{C}^{N_{\text{r}}N_{\text{t}} \times N_{\text{r}}N_{\text{t}}}$ representing the spatial correlation matrix. The message from the observed factor node $f$ to the variable node is represented as
\begin{align}
       {m_{f \to {\mathbf{h}}}} &= \exp \left\{ {\mathbb{E}\left\{ {\ln \left( {p\left( {{\mathbf{y}}|{\mathbf{s}},{\mathbf{h}}} \right)} \right)} \right\}} \right\} \notag \\
       &\propto \exp \left\{ {\mathbb{E}\left\{ { - \frac{1}{{\sigma _w^2}} {{{\left\| {{\mathbf{y}} - {\mathbf{Sh}}} \right\|}^2}}  } \right\}} \right\}.\label{equ_m_f2v}
\end{align}

The expectation $\mathbb{E}\{  \cdot  \}$ in \eqref{equ_m_f2v} requires prior knowledge of the transmitted symbols, which are acquired via soft remapping from \eqref{equ_DEC_dhat} in the previous JCDD iteration. Specifically, the symbol estimates ${{\widetilde{\mathbf{s}}}_n} \in \mathbb{C}^{W \times 1}$ are calculated as
\begin{equation}
       \label{equ_Stilde}
       {{\widetilde{\mathbf{s}}}_n} = \sqrt \rho \, {{\mathbf{p}}_n} + \sqrt {1 - \rho } \, {{\widehat{\mathbf{d}}}_n},
\end{equation}
and reshaped into a matrix ${{\widetilde{\mathbf{S}}}_n} \in \mathbb{C}^{K \times T}$. With ${{\widehat{\mathbf{s}}}_n}$ denoting the $t$-th column of ${{\widetilde{\mathbf{S}}}_n}$, the mean of ${\mathbf{S}}$ is given by
\begin{equation}
       \label{equ_Shat}
       {\widehat{\mathbf{S}}} = \mathbb{E}\left\{ {\mathbf{S}} \right\} = {{\mathbf{I}}_{{N_{\text{r}}}}} \otimes \left( {\left( {{{\mathbf{1}}_{1 \times {N_{\text{t}}}}} \otimes {{\mathbf{I}}_K}} \right){\text{Diag}}\left( {{\widehat{\mathbf{s}}}} \right)} \right),
\end{equation}
where ${\widehat{\mathbf{s}}} = {[ {{{\widehat{\mathbf{s}}}_1^{\mathrm{T}}}, \ldots, {{\widehat{\mathbf{s}}}_{{N_{\text{t}}}}^{\mathrm{T}}}} ]^{\mathrm{T}}}$ is the concatenation of the estimated transmitted symbols.

Moreover, the variance of data estimates can be computed according to \eqref{equ_DEC_dhat} as 
\begin{equation}
       \label{equ_vtilde}
       {{\widetilde{v}}_{n}^{w}} = \sum\limits_{a \in \mathcal{A}} {{{\left( {a - {{\widehat{d}}_{n}^{w}}} \right)}^2} \cdot {\text{Pr}}\left( {a|{L}_{{\text{A}},n}^{w}} \right)},
\end{equation}
and ${{\widetilde{\mathbf{v}}}_{n}}=[{{\widetilde{v}}_{n}^{w}}] \in \mathbb{C}^{W \times 1}$  is reshaped into a matrix ${{\widetilde{\mathbf{V}}}_n} \in \mathbb{C}^{K \times T}$. With ${{{\mathbf{v}}}_n}$ denoting the $t$-th column of ${{\widetilde{\mathbf{V}}}_n}$, the covariance of ${\mathbf{S}}$ is expressed as
\begin{equation}
       \label{equ_Sigma_Shat}
       {{\widehat{\boldsymbol{\Sigma}}}_{\mathbf{S}}} = \mathbb{E}\left\{ {{{\mathbf{S}}^{\mathrm{H}}}{\mathbf{S}}} \right\} - {{\widehat{\mathbf{S}}}^{\mathrm{H}}}{\widehat{\mathbf{S}}} = \left( {1 - \rho }  \right)\left( {{{\mathbf{I}}_{{N_{\text{r}}}}} \otimes {\text{Diag}}\left( {\mathbf{v}} \right)} \right),
\end{equation}   
where ${{\mathbf{v}}} = {[ {{{{\mathbf{v}}}_1^{\mathrm{T}}},  \ldots, {{{\mathbf{v}}}_{{N_{\text{t}}}}^{\mathrm{T}}}} ]^{\mathrm{T}}}$. Therefore, the message ${m_{f \to {\mathbf{h}}}}$ in \eqref{equ_m_f2v} becomes 
\begin{equation}
       \label{equ_m_f2v_2}
       {m_{f \to {\mathbf{h}}}} \propto \exp \left\{ { - \frac{1}{{\sigma _w^2}}\left( {{{ \| {{\mathbf{y}} - {\widehat{\mathbf{S}}\mathbf{h}}} \|}^2} + {{\mathbf{h}}^{\mathrm{H}}}{{\widehat{\boldsymbol{\Sigma}}}_{\mathbf{S}}}{\mathbf{h}}} \right)} \right\}.
\end{equation}

Finally, the VMP-based channel estimation is summarized as
\begin{subequations}
       \label{equ_hhat_vmp}
       \begin{align}
         \widehat {\boldsymbol{\Sigma}}  &= {{{{\widehat{\mathbf{S}}}^{\mathrm{H}}}{\widehat{\mathbf{S}}} + {{\widehat{\boldsymbol{\Sigma}}}_{\mathbf{S}}} + \sigma _w^2 {\boldsymbol{\Sigma}}_{{\text{p}}}^{ - 1}}}, \\
         {\widehat{\mathbf{h}}} &= {\widehat{\boldsymbol{\Sigma}}}^{-1} {{\widehat{\mathbf{S}}}^{\mathrm{H}}}{\mathbf{y}}.
       \end{align}
\end{subequations}
As shown in \eqref{equ_hhat_vmp}, time-correlation properties are not utilized in the derived VMP-based refined channel estimation. This effectively addresses the performance loss caused by mismatched channel characteristics. 

\subsection{VMP-L}
\label{sec_Adaptive_VMPL} 
While the VMP-based method demonstrates strong generalization capability, its implementation through spatial-frequency domain concatenation, as shown in \eqref{equ_y_t}, results in high-dimensional matrix operations and significantly increased computational complexity. To address this, we derive a simplified variational inference method, termed VMP-L, which provides adaptive estimation with substantially lower complexity.

\begin{figure}[t]
  \centering
  \begin{minipage}{2.8in}
    \centerline{\includegraphics[width=2.8in]{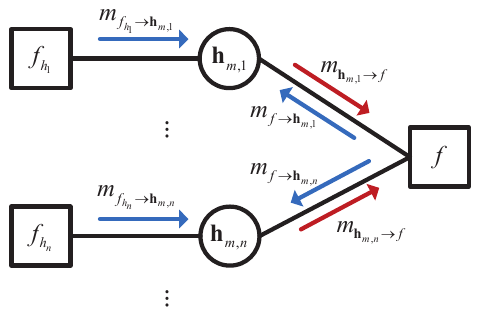}}
  \end{minipage}
  \caption{The channel vector factor graph representation for VMP-L, where the channel coefficients corresponding to the $m$-th receive antenna at the $t$-th time instant are represented by $N_{\text{t}}$ variable nodes.}
  \label{fig_vmpL_factor}
\end{figure}

Specifically, the channel prior is factorized with respect to the $n$-th transmit antenna, allowing the channel coefficients to be represented by $N_{\text{t}}$ variable nodes, where ${{\mathbf{h}}_{m,n}^{(t)}}$ denotes the $n$-th variable node. For notational simplicity, the superscript ${(  \cdot  )^{( t )}}$ is omitted, and the corresponding factor graph is shown in \figref{fig_vmpL_factor}.

Following \eqref{equ_y_m_t}, the message from the observed factor node $f$ to variable node ${{\mathbf{h}}_{m,n}}$ is expressed in \eqref{equ_m_f2v_vmpl}, 
\begin{figure*}[ht]
       \centering
       \begin{equation}
         \label{equ_m_f2v_vmpl}
         {m_{f \to {{\mathbf{h}}_{m,n}}}} \propto \exp \left\{ { - \frac{1}{{\sigma _w^2}}\left( {{{\Big\| {{{\mathbf{y}}_m} - \sum_{n' \ne n} {{{\text{Diag}} \left({{\widehat{\mathbf{s}}}_{n'}}\right)} \cdot {{\widehat{\mathbf{h}}}_{m,n'}}}  - {{\text{Diag}}\left({{\widehat{\mathbf{s}}}_n}\right)}\cdot{{\mathbf{h}}_{m,n}}} \Big\|}^2}{+}{\mathbf{h}}_{m,n}^{\mathrm{H}}{{\widehat{\boldsymbol{\Sigma}}}_{{\mathbf{s}}_n}}{{\mathbf{h}}_{m,n}}} \right)} \right\}
       \end{equation}
       \vspace*{1ex}
       \begin{equation}
         \label{equ_m_v2f_vmpl}
         {m_{{{\mathbf{h}}_{m,n}} \to f}} = {m_{f \to {{\mathbf{h}}_{m,n}}}} \cdot {m_{{f_{h_n}} \to {{\mathbf{h}}_{m,n}}}}
         \propto \exp \left\{ { - {{\left( {{\mathbf{h}}_{m,n} - {\widehat{\mathbf{h}}}_{m,n}} \right)}^{\mathrm{H}}}{{\widehat {\boldsymbol{\Sigma}} }_{n}^{ - 1}}\left( {{\mathbf{h}}_{m,n} - {\widehat{\mathbf{h}}}_{m,n}} \right)} \right\}
       \end{equation}
       \vspace*{1ex}
       \begin{equation}
         \label{equ_CE_ls_refined_noise}
         {{{\mathbf{n}}_{{\text{eff}}}}} = {\sum_{n' \ne n} {{\mathbf{P}}_n^{ - 1}{{\mathbf{P}}_{n'}}{{\mathbf{\Delta h}}_{m,n'}}}  
         + {{\left( {\sqrt \rho  {{\mathbf{P}}_n}} \right)}^{ - 1}}{{\mathbf{w}}_m}}
         + \sqrt {\frac{{1 - \rho }}{\rho }} \sum\limits_{n' = 1}^{{N_{\text{t}}}} {{\mathbf{P}}_n^{ - 1}\left( {{\widehat{\mathbf{D}}}_{n'}{{\mathbf{\Delta h}}_{m,n'}} + {{\mathbf{\Delta D}}_{n'}}{\widehat{\mathbf{h}}}_{m,n'} + {{\mathbf{\Delta D}}_n'}{{\mathbf{\Delta h}}_{m,n'}}} \right)}  
       \end{equation}
       \hrulefill
\end{figure*}
where the covariance of the estimated data symbols is ${{\widehat{\boldsymbol{\Sigma}}}_{{\mathbf{s}}_n}}=(1-\rho){\text{Diag}}( {\mathbf{v}_n} )$. The prior $f_{h_n}$ is given by
\begin{equation}
       \label{equ_m_hp2v_vmpl}
       {m_{{f_{h_n}} \to {{\mathbf{h}}_{m,n}}}} \propto \exp \left\{ { - {{ {{\mathbf{h}}_{m,n}^{\mathrm{H}}} }}{\left({{\mathbf{R}}_{{\mathbf{hh}}}^{{\text{Freq}}}}\right)}^{ - 1} {{\mathbf{h}}_{m,n}} } \right\}.
\end{equation}
The outgoing message from the variable node to the observation factor, ${m_{\mathbf{h}_{m,n} \to f}}$, is given in \eqref{equ_m_v2f_vmpl}. The sequential variational updates are summarized as follows:\footnote{The proposed VMP-L approach decouples the probabilistic model and leverages message passing to formulate the inference problem based solely on frequency-domain correlation. This design effectively avoids the high-dimensional computations required in VMP. A detailed complexity analysis is provided in \secref{sec_simulation}-D.}
\begin{subequations}
       \label{equ_hhat_vmpl}
       \begin{align}
         \label{equ_yhat_vmpl}
         {{\widetilde{\mathbf{y}}}_{m,n}} &= {{\mathbf{y}}_m} - \sum_{n' \ne n} {{{\text{Diag}} \left({{\widehat{\mathbf{s}}}_{n'}}\right)} \cdot{{\widehat{\mathbf{h}}}_{m,n'}}}, \\
         {\widehat {\boldsymbol{\Sigma}}}_n  &= { {{{{\text{Diag}}\left({{\widehat{\mathbf{s}}}_n}{{\widehat{\mathbf{s}}}_n^{\mathrm{H}}}\right)}} + {{\widehat{\boldsymbol{\Sigma}}}_{{\mathbf{s}}_n}} + \sigma _w^2 {\left({{\mathbf{R}}_{{\mathbf{hh}}}^{{\text{Freq}}}}\right)}^{ - 1}} }, \\
         {\widehat{\mathbf{h}}} &= {\widehat{\boldsymbol{\Sigma}}}_n^{-1} {{\text{Diag}}\left({{\widehat{\mathbf{s}}}_n}\right)^{\mathrm{H}}}{{\widetilde{\mathbf{y}}}_{m,n}}.
       \end{align}
\end{subequations}
Note that each time a new message ${m_{{{\mathbf{h}}_{m,n}} \to f}}$ is computed, the observation in \eqref{equ_yhat_vmpl} is recalculated using the latest prior estimate ${\widehat{\mathbf{h}}}_{m,n}$.

\subsection{DL}
\label{sec_Adaptive_DL}
\begin{figure*}[t]
  \centering
  \includegraphics[width=1.0\textwidth]{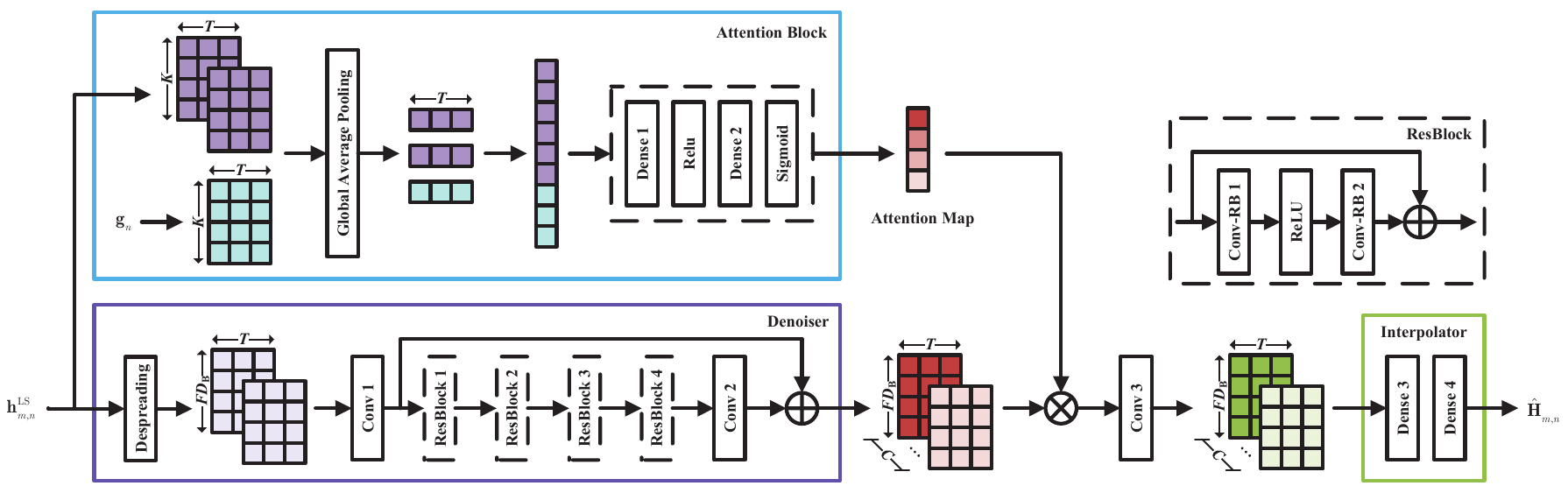}
  \caption{Architecture of the proposed DL-based refined channel estimator.}
  \label{fig_dl}
\end{figure*}

The adverse effects of time-domain statistical mismatch can also be mitigated using a DL approach. As illustrated in \figref{fig_dl}, the proposed DL-based channel estimation network consists of three modules: a denoiser, an attention-enhanced feature reweighting block, and an interpolator.

\subsubsection{Denoiser}
A residual learning-based CNN is developed to enhance low-quality LS-based channel estimates into high-resolution reconstructions. In conventional systems, the network input consists of sparse estimates located at pilot positions \cite{DLDenoiseLi}. In contrast, SIP systems involve superimposed pilot-data transmission across all time-frequency REs, resulting in dense, high-dimensional input tensors. This full-grid occupation not only leads to excessive training overhead but also complicates feature extraction for convolutional filters.

To address this, a despreading operation is introduced prior to CNN-based denoising.
The initial LS-based estimates, derived in \eqref{equ_CE_ls_refined}, are expressed as 
\begin{equation}
       \label{equ_ls_dl}
       {\mathbf{h}}_{m,n}^{{\text{LS}}} = {{\mathbf{h}}_{m,n}} + {{{\mathbf{n}}_{{\text{eff}}}}},
\end{equation}
where the residual term ${{{\mathbf{n}}_{{\text{eff}}}}}$ is decomposed in \eqref{equ_CE_ls_refined_noise} into components reflecting estimation errors in both the data symbols ${{\mathbf{\Delta D}}_n}$ and channel coefficients ${{\mathbf{\Delta h}}_{m,n}}$. Exploiting the zero-mean properties of these errors and the additive noise, the despreading operation \cite{DesCite, SJCDDQian} performs averaging over $L_1$ and $L_2$ neighboring REs in the frequency and time domains, respectively. This process suppresses residual interference and enhances feature consistency. As shown in \figref{fig_dl}, the coarse estimates output by the despreading operator are concatenated into real and imaginary parts to form two feature maps of size ${FD}_{\text{B}} \times T$, where ${FD}_{\text{B}}=K/{L_1}$.

In the denoiser module, noise suppression is achieved using a residual CNN architecture inspired by \cite{DLDenoiseLi, DLDenoiseLi2}.
Initially, the input features are convolved with $C$ filters of size $F \times F \times 2$ in the Conv1 layer. This is followed by four residual blocks (ResBlocks), where each block contains two convolutional layers separated by a ReLU activation, and each layer uses $C$ filters of size $F \times F \times C$. Finally, the denoised feature maps are refined through Conv2, which applies an additional $C$ convolutional filters.

\begin{figure*}[ht]
  \centering
  \begin{minipage}{0.24\linewidth}
    \centerline{\includegraphics[scale=0.7]{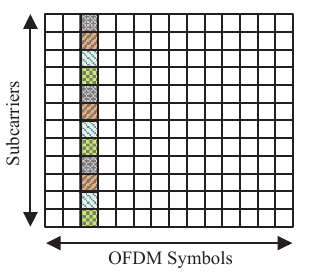}}
    \centerline{\small{(a) 1P}}
  \end{minipage}
  \begin{minipage}{0.32\linewidth}
    \centerline{\includegraphics[scale=0.7]{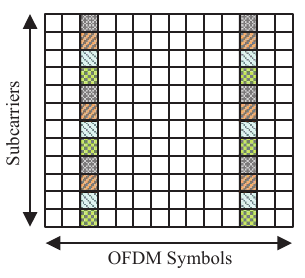}}
    \centerline{\small{(b) 2P}}
  \end{minipage}
  \begin{minipage}{0.36\linewidth}
    \centerline{\includegraphics[scale=0.7]{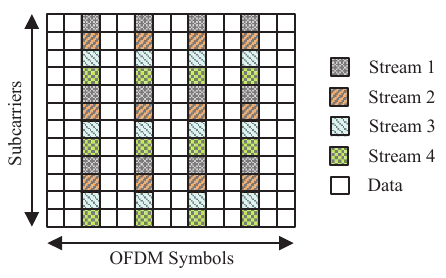}}
    \leftline{\hspace{0.76in} \small{(c) 4P}}
  \end{minipage}
  \caption{Illustration of the pilot pattern used in the OP transmission scheme.}
  \label{fig_op_pilot}
  \vspace{-1em}
\end{figure*}

\subsubsection{Attention Block}
Although the despreading operation in the denoiser helps suppress noise, it may inadvertently discard useful time-frequency correlation information. This can reduce the CNN's robustness under mismatched channel conditions. To address this, an attention mechanism \cite{DLAttYang, DLAttGao, DLAttGao2} is integrated into the network to better capture the statistical structure of time-varying channels.
As shown in \figref{fig_dl}, the attention block assigns different weights to the features output by the denoiser, thereby emphasizing those most relevant to the current channel distribution. This feature importance weighting improves the effectiveness of subsequent feature fusion in the convolution layer. 

To guide the learning of attention weights, the LS-based channel estimates ${\mathbf{h}}_{m,n}^{{\text{LS}}}$ are provided as input to the attention block, allowing extraction of intrinsic correlation structures across time and frequency. To mitigate error propagation caused by inaccurate interference cancellation in \eqref{equ_CE_ls_refined_noise}, a self-supervised mechanism is incorporated. This mechanism quantifies the reliability of the soft data estimates and embeds this confidence information \cite{SupervisedFin} into the attention module. Specifically, based on \eqref{equ_DEC_dhat}, the logits corresponding to the soft symbol ${{\widehat{d}}_{n}^{w}}$ from the previous JCDD iteration are used to compute the confidence score: 
\begin{subequations}
       \label{equ_dl_confidence}
       \begin{align}
         {g_{n}^{w}} &\triangleq \left| {{l_{{i^ * }}} - {l_{{i^{ *  * }}}}} \right|, \\
         {l_{{i^ * }}} &\triangleq \mathop{\arg\max}\limits_i \, {l_i}, ~~
         {l_{{i^{ *  * }}}} \triangleq \mathop{\arg\max}\limits_{i \ne {i^ * }} \, {l_i},
       \end{align}
\end{subequations}
where a higher value of ${g_{n}^{w}}$ indicates more reliable feedback data.

The global information, comprising the real and imaginary parts of ${\mathbf{h}}_{m,n}^{{\text{LS}}}$ and the confidence feature vector ${{\mathbf{g}_n}} = [{g_{n}^{w}}] \in {\mathbb{R}^{W \times 1}}$, is aggregated via global average pooling into a squeezed feature $\mathbf{f} \in \mathbb{R}^{1\times 1 \times 3T}$. Based on $\mathbf{f}$, the attention map $\mathbf{a} \in \mathbb{R}^{1\times 1 \times C}$ is generated by a two-layer deep neural network (DNN). The first dense layer with $N_{h1}$ neurons is followed by a ReLU activation, and the second layer with $C$ neurons is followed by a Sigmoid activation. The resulting attention weights are applied in a channel-wise manner to reweight the feature maps of size ${FD}_{\text{B}} \times T \times C$, which are then passed through $C$ filters in Conv3 for further processing.

\subsubsection{Interpolator}
Finally, frequency-domain interpolation is performed to recover the channel estimates over the full set of time-frequency REs.
A two-layer DNN serves as the interpolator, processing input features of size ${FD}_{\text{B}} \times T \times C$. These features are reshaped into a $T \times (C \cdot {FD}_{\text{B}})$ matrix, denoted as ${{\widetilde{\mathbf{H}}}_{m,n}}$, before being passed through the DNN with hidden layer sizes $\{N_{h2}, 2K\}$ neurons. The output has dimensions $T \times 2K$ and corresponds to the real and imaginary parts of the estimated channel ${\widehat{\mathbf{h}}}_{m,n}$.

The proposed DL network is trained in a supervised manner by minimizing the mean squared error between the estimated and ground-truth channel coefficients, defined as
\begin{equation}
       \label{equ_dl_loss}
       {\text{loss}} = \frac{1}{{{N_{\text{r}}}{N_{\text{t}}}W}} {\sum\limits_{m = 1}^{{N_{\text{r}}}} {\sum\limits_{n = 1}^{{N_{\text{t}}}} {{{\left\| {{{\widehat{\mathbf{h}}}_{m,n}} - {{\mathbf{h}}_{m,n}}} \right\|}^2}} } } .
\end{equation}
The despreading operation uses average lengths $L_1 = 6$ and $L_2 = 2$ in the frequency and time domains, respectively. The convolutional layers are configured with $C = 8$ channels and a kernel size $F = 3$. The dense layers have $N_{h1} = 16$ and $N_{h2} = 128$ neurons. The model is optimized using the Adam optimizer with a learning rate of 0.001 and trained for 100 epochs with a batch size of 128.

\section{Simulation Results}
\label{sec_simulation}
In this section, we present numerical results. First, the simulation parameters for the MIMO-OFDM system are introduced. We then compare the performance of the proposed JCDD structure for the SIP receiver with that of the traditional OP receiver. Furthermore, we assess the generalization capability of the proposed adaptive channel estimation schemes under mismatched channel conditions and provide an analysis of computational complexity. 

\vspace{-1em}
\subsection{Parameter Settings}
An uplink MIMO-OFDM system is simulated with $N_{\text{r}}$ receive antennas at the BS and $N_{\text{t}}$ transmit antennas at the user equipment (UE). Two antenna configurations are considered: $(N_{\text{r}}, N_{\text{t}}) = (2, 2)$ and $(64, 4)$. The system is configured with $K = 144$ subcarriers and $T = 14$ OFDM symbols, corresponding to 12 resource blocks (RBs). In the SIP transmission scheme, all time-frequency REs are allocated for data transmission. An LDPC encoder maps $k = 4032$ data bits to coded bits at a rate of $r = 1/2$ for each transmit antenna, and the resulting codewords are modulated using 16-QAM. Pilot symbols are superimposed on all REs with a power allocation ratio of $\rho = 0.3$ \cite{SJCDDJing,SJCDDQian}.

\begin{table}[t]
  \caption{Parameters for the Simulation}
  \label{tab_mimo_ofdm}
  \centering
  \begin{tabular}{ll}
    \toprule
    Parameter & Value \\
    \midrule
    Antennas & $2 \times 2$, $64 \times 4$ \\
    Time-frequency REs & $K=144,~T=14$ \\
    Power Delay Profile & CDL-C \\
    Carrier Frequency & 3.5 GHz \\
    Subcarrier Spacing & 30 kHz \\
    Delay Spread & 200 ns \\
    Testing UE Speed & $v \in \{15,324\}$ km/h \\
    Modulation & 16-QAM \\
    Channel Coding Scheme & LDPC \\
    Code Rate & $r=1/2$ \\
    Power Allocation & $\rho=0.3$ \\
    Detector & LMMSE \\
    \bottomrule
  \end{tabular}
\end{table}

Channel propagation is modeled using the clustered delay line (CDL) channel model with the CDL-C power delay profile \cite{CDLCite}. Both low- and high-mobility scenarios are evaluated, corresponding to UE speeds of $v = 15$ km/h and $v = 324$ km/h, respectively. Channel correlation matrices are computed as the second-order statistical average from $10^5$ MIMO-OFDM channel samples. All simulations are implemented using the NVIDIA Sionna package \cite{SionnaHoy}. The key system parameters are summarized in \tabref{tab_mimo_ofdm}.

\begin{figure*}[t]
  \centering
  \begin{minipage}{0.45\linewidth}
    \centerline{\includegraphics[width=3.5in]{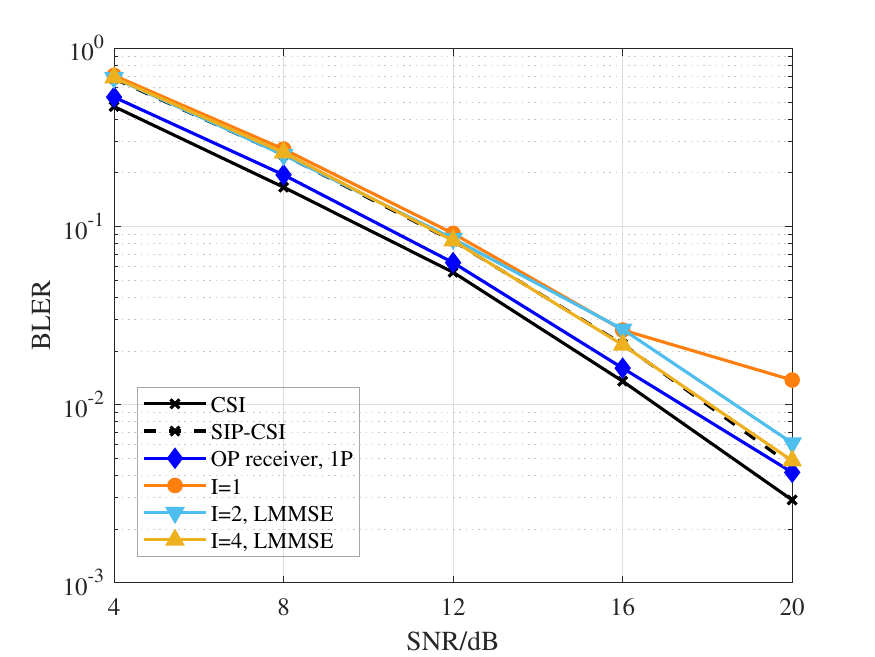}}
    \centerline{\small{(a) BLER at low speed.}}
  \end{minipage}\quad\quad\quad
  \begin{minipage}{0.45\linewidth}
    \centerline{\includegraphics[width=3.5in]{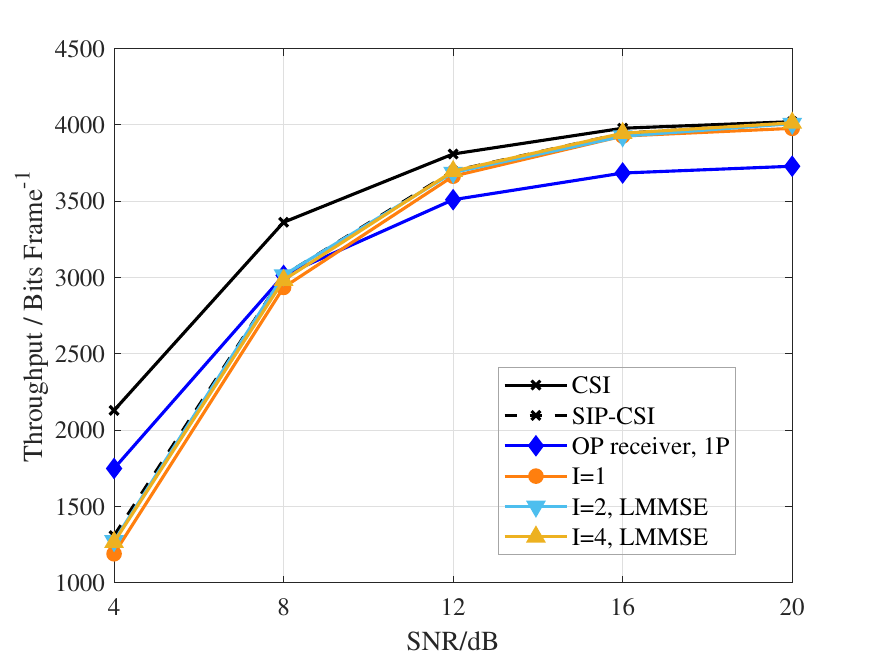}}
    \centerline{\small{(b) Throughput at low speed.}}
  \end{minipage}
  \vspace{0.1\baselineskip}
  \begin{minipage}{0.45\linewidth}
    \centerline{\includegraphics[width=3.5in]{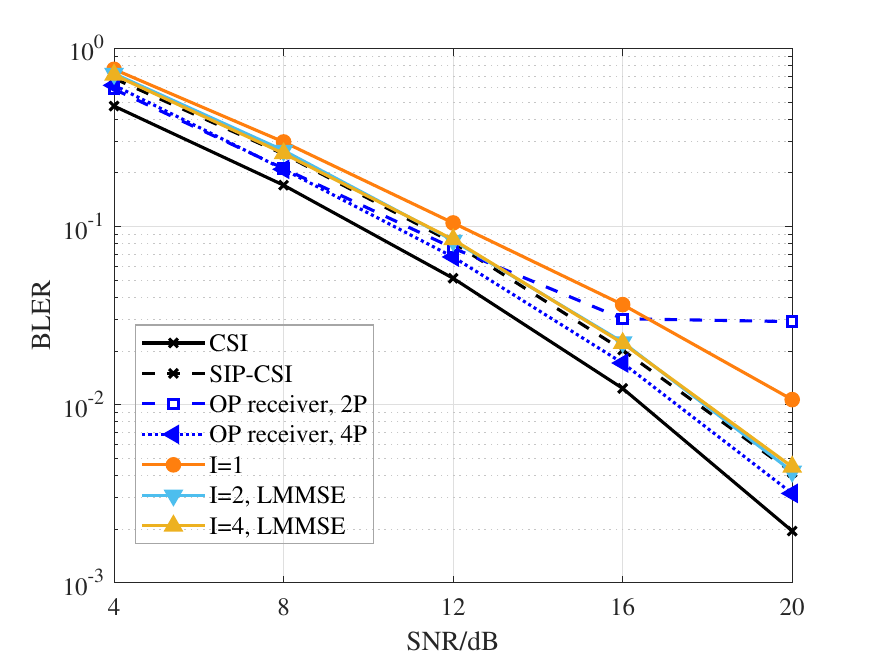}}
    \centerline{\small{(c) BLER at high speed.}}
  \end{minipage}\quad\quad\quad
  \begin{minipage}{0.45\linewidth}
    \centerline{\includegraphics[width=3.5in]{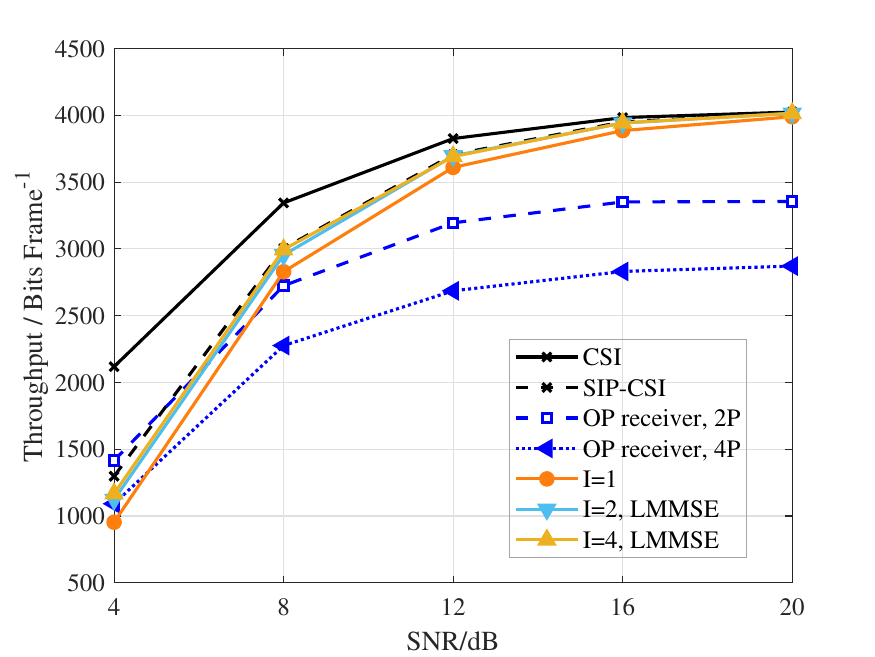}}
    \centerline{\small{(d) Throughput at high speed.}}
  \end{minipage}
  \vspace{0.1\baselineskip}
  \caption{Performance comparisons between the OP receiver and our proposed iterative receiver in the $2\times2$ MIMO-OFDM system.}
  \label{fig_comp_op_N2}
  \vspace{-1em}
\end{figure*}

\vspace{-1em}
\subsection{Comparisons with OP Receiver}
We begin by analyzing the performance of the proposed iterative receiver across varying numbers of JCDD iterations, specifically $I \in \{1, 2, 4\}$. The curves labeled ``$I=2$, LMMSE'' and ``$I=4$, LMMSE''  utilize refined channel estimates obtained via LMMSE,\footnote{Note that matched channel statistics are considered in this subsection, under which both the proposed VMP-based and DL-based enhancements exhibit performance comparable to the LMMSE-based method, as evidenced in \figref{fig_comp_mm}. Therefore, to streamline the analysis, we focus on LMMSE-based refined channel estimation for direct comparison with the OP baselines. Detailed evaluations of generalization are deferred to \secref{sec_simulation}-C.} as outlined in \eqref{equ_CE_lmmse_refined}. In this subsection, we assess both the block error rate (BLER) and throughput. The throughput is defined as the number of correctly received bits per frame and is computed as
\begin{equation}
       \label{equ_throughput}
       R = K \times T \times \Omega \times r \times Q \times \left( {1 - {\text{BLER}}} \right) ,
\end{equation}
where $\Omega$ denotes the proportion of REs allocated to data symbols.

The proposed SIP receiver is benchmarked against a conventional OP-based receiver that uses LMMSE-based channel estimation and detection. Three pilot patterns standardized in 5G NR are considered. The ``1P'' pattern, shown in \figref{fig_op_pilot}(a), is applied in low-mobility scenarios, while the ``2P'' and ``4P'' patterns are used to accommodate high-mobility environments due to their improved temporal tracking capabilities. In OP-based transmission, each transmit antenna is assigned $K/N_{\text{t}}$ dedicated subcarriers, and the case with $N_{\text{t}} = 4$ is illustrated in \figref{fig_op_pilot}. Additionally, two ideal benchmarks are provided assuming perfect CSI: one for conventional data transmission without pilot superposition ($\rho = 0$), labeled ``CSI,'' and one for SIP with pilot power ratio $\rho = 0.3$, labeled ``SIP-CSI.''

Performance evaluations for the $2 \times 2$ MIMO-OFDM system are presented in \figref{fig_comp_op_N2}. In \figref{fig_comp_op_N2}(a), it is observed that under low-speed conditions, channel variations are sufficiently small such that OP transmission with pilots placed on a single OFDM symbol enables accurate LMMSE-based interpolation. Consequently, the performance gap between the ``CSI'' upper bound and the ``OP receiver, 1P'' configuration is negligible. Additionally, the proposed JCDD receiver with $I = 2$ iterations achieves BLER performance close to the SIP-CSI bound, and further increasing the number of iterations to $I = 4$ results in minor improvement. While the OP receiver exhibits a slight edge in estimation accuracy, it suffers from reduced spectral efficiency due to pilot overhead. This trade-off is evident in \figref{fig_comp_op_N2}(b), where the proposed JCDD receiver with $I = 2$ outperforms the OP receiver by 7.5\% in throughput.

We further evaluate system performance under high-mobility conditions. As illustrated in \figref{fig_comp_op_N2}(c), the proposed JCDD receiver converges after $I = 2$ iterations, achieving BLER performance comparable to the ideal ``SIP-CSI'' bound. In contrast, the ``2P'' pilot pattern employed by OP receiver fails to capture rapid channel variations, resulting in noticeable performance degradation at high SNRs. While the ``4P'' pilot pattern provides improved channel estimates, it incurs significant overhead with $\Omega = 10/14$. \figref{fig_comp_op_N2}(d) shows that the proposed JCDD receiver with $I = 2$ achieves a 19.7\% throughput gain over the ``OP receiver, 2P,'' and a 39.8\% gain over the ``OP receiver, 4P.''

\begin{figure*}[t]
  \centering
  \begin{minipage}{0.45\linewidth}
    \centerline{\includegraphics[width=3.5in]{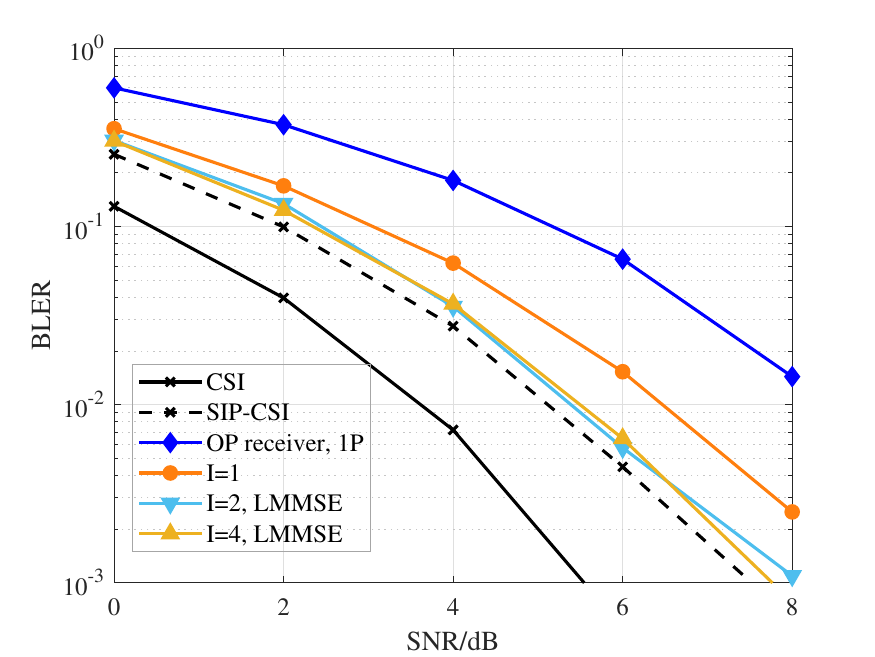}}
    \centerline{\small{(a) BLER at low speed.}}
  \end{minipage}\quad\quad\quad
  \begin{minipage}{0.45\linewidth}
    \centerline{\includegraphics[width=3.5in]{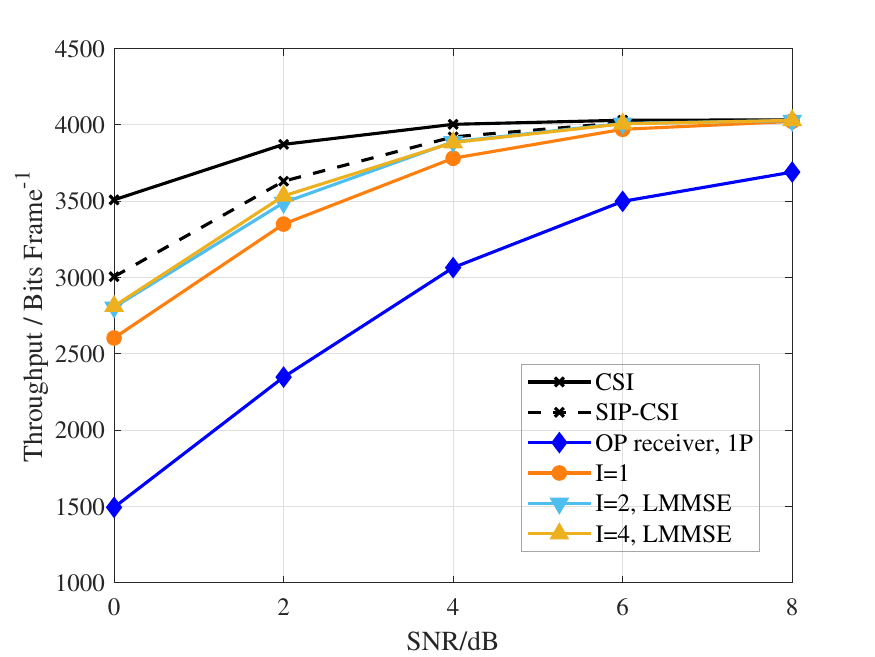}}
    \centerline{\small{(b) Throughput at low speed.}}
  \end{minipage}
  \vspace{0.1\baselineskip}
  \begin{minipage}{0.45\linewidth}
    \centerline{\includegraphics[width=3.5in]{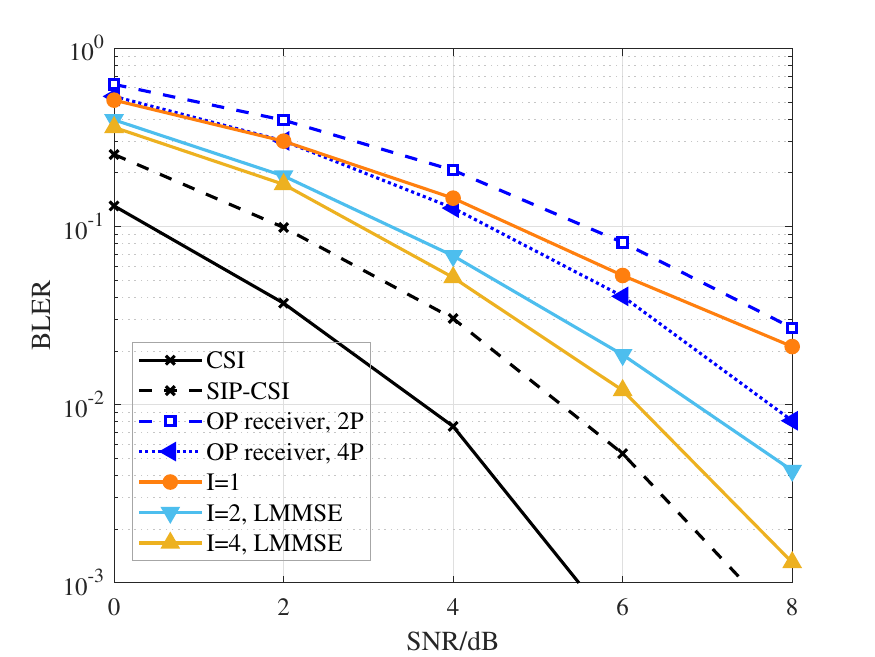}}
    \centerline{\small{(c) BLER at high speed.}}
  \end{minipage}\quad\quad\quad
  \begin{minipage}{0.45\linewidth}
    \centerline{\includegraphics[width=3.5in]{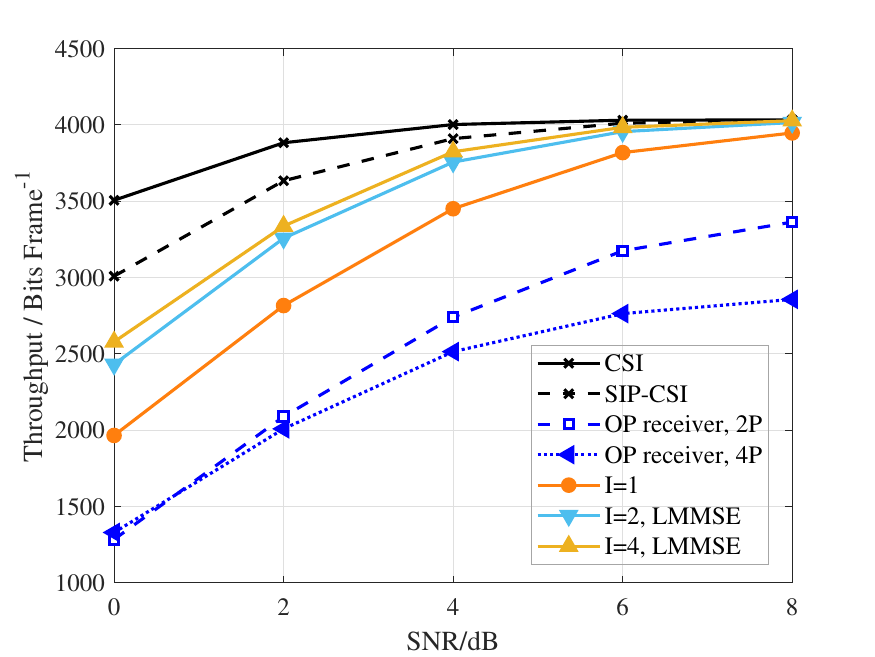}}
    \centerline{\small{(d) Throughput at high speed.}}
  \end{minipage}
  \vspace{0.1\baselineskip}
  \caption{Performance comparisons between the OP receiver and our proposed iterative receiver in the $64\times4$ MIMO-OFDM system.}
  \label{fig_comp_op_N4}
  \vspace{-1em}
\end{figure*}

Subsequently, performance evaluations for the $64 \times 4$ MIMO-OFDM system are presented in \figref{fig_comp_op_N4}. As shown in \figref{fig_comp_op_N4}(a), under low-mobility conditions, the proposed JCDD receiver with $I = 2$ consistently yields significant performance gains, approaching the upper bound for SIP transmission. Notably, the OP receiver using the ``1P'' pilot pattern performs poorly, despite the quasi-static nature of the channel. This degradation is attributed to the limited pilot allocation in the frequency domain, as each transmit antenna is assigned only $K/N_{\text{t}}$ subcarriers, which restricts channel estimation accuracy. In contrast, the SIP scheme offers a key advantage by superimposing pilot symbols across all REs, leaving the pilot allocation independent of the number of spatial streams. Additionally, as shown in \figref{fig_comp_op_N4}(b), the JCDD receiver with $I = 2$ achieves a 9.1\% throughput improvement over the OP receiver.

\figref{fig_comp_op_N4}(c) presents the detection accuracy in the $64 \times 4$ MIMO-OFDM system under high-mobility conditions. The OP receivers exhibit limited BLER performance due to insufficient pilot density in the frequency domain, irrespective of the number of OFDM symbols allocated to pilot transmission. The proposed JCDD receiver with $I = 4$ iterations provides an approximate 1 dB improvement over the $I = 2$ configuration. However, this performance gain is accompanied by nearly doubling the processing time, and according to \figref{fig_comp_op_N4}(d), the throughput benefit of $I = 4$ over $I = 2$ is marginal. The JCDD receiver with $I = 2$ achieves notable throughput improvements of 19.4\% and 40.5\% relative to the OP receivers with ``2P'' and ``4P'' pilot patterns, respectively.

To summarize, the proposed SIP receiver offers significant performance benefits compared to the traditional OP receiver. By superimposing pilot symbols over all time-frequency REs, the SIP scheme enables more accurate tracking of channel variations, making it particularly effective for systems with multiple transmit antennas and rapid channel dynamics. Furthermore, since no additional pilot overhead is introduced, the SIP scheme achieves substantial throughput improvements over OP-based designs. The setting of $I = 2$ iterations strikes a favorable balance between performance and computational efficiency, and is therefore adopted in the remaining performance evaluations.

\begin{table*}[!t]
  \caption{Complexity Analysis}
  \label{tab_complexity}
  \centering
  \begin{tabular}{llcc}
    \toprule
    \multirow{2}{*}{Algorithm} & \multirow{2}{*}{Complexity} & \multicolumn{2}{c}{Runtime (ms)} \\
    & & (a)~$2\times2$ MIMO & (b)~$64\times4$ MIMO \\
    \midrule
    LMMSE & $\mathcal{O}( {{N_{\text{r}}}{N_{\text{t}}}( {{K^3}T + {T^3}K} )} )$ & 0.57 & 4.89 \\
    VMP-L & $\mathcal{O}\left( {T{N_{\text{t}}}\left( {{N_{\text{r}}} + 1} \right){K^3}} \right)$ & 0.53 & 5.05 \\
    VMP & $\mathcal{O}( {T{{( {{N_{\text{r}}}{N_{\text{t}}}K} )}^3}} )$ & 38.79 & -- \\
    DL & $\mathcal{O}\left( {{N_{\text{r}}}{N_{\text{t}}}T\left( {{FD}_{{\text{B}}} \cdot 10{F^2}{C^2} + {N_{h2}}\left( {F{D_{\text{B}}} \cdot C + {2K}} \right)} \right)} \right)$ & 1.33 & 6.71 \\
    \bottomrule
  \end{tabular}
  \vspace{-1em}
\end{table*}

\begin{figure}[t]
  \centering
  \begin{minipage}{3.5in}
    \centerline{\includegraphics[width=3.5in]{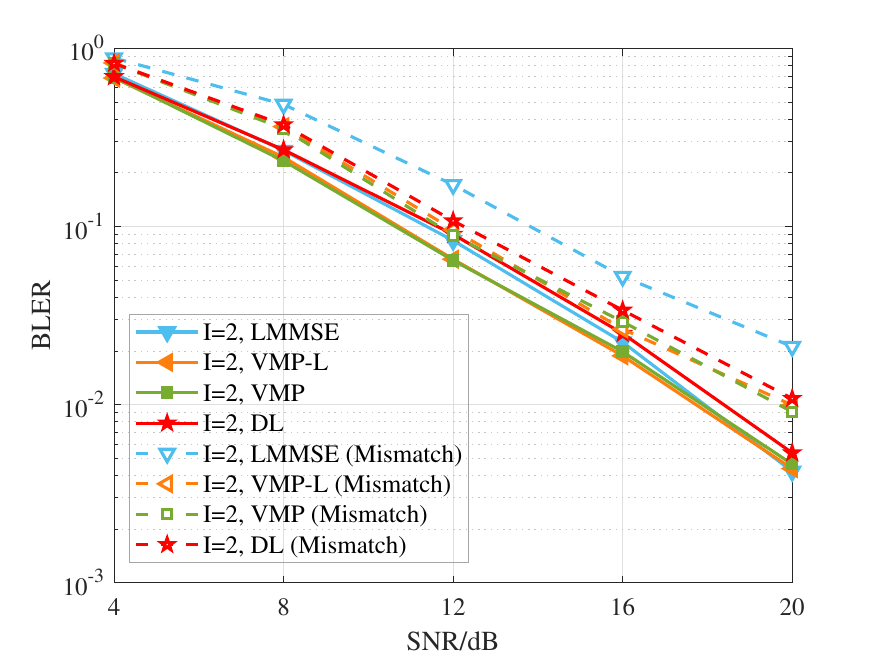}}
    \centerline{\small{(a) $2\times2$ MIMO}}
  \end{minipage}
  \hfill
  \begin{minipage}{3.5in}
    \centerline{\includegraphics[width=3.5in]{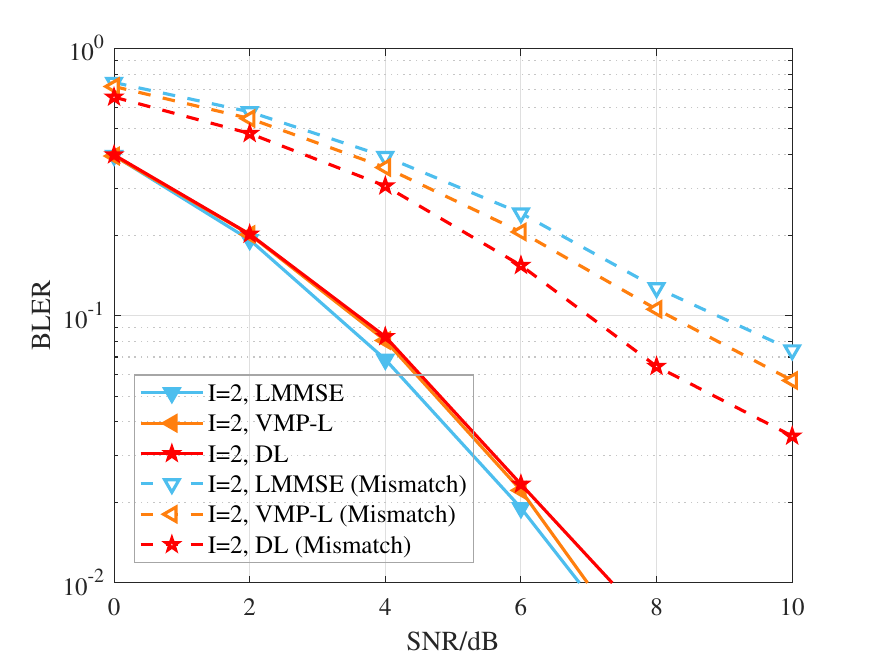}}
    \centerline{\small{(b) $64\times4$ MIMO}}
  \end{minipage}
  \caption{Performance comparisons of the proposed JCDD designs under both matched and mismatched cases.}
  \label{fig_comp_mm}
\end{figure}

\vspace{-1em}
\subsection{Adaptation to Mismatched Channel Statistics}
In this subsection, the BLER performance of the proposed refined channel estimation schemes is presented, focusing on a high-mobility scenario with a UE speed of $v = 324$ km/h. Both matched and mismatched channel conditions are analyzed. In the matched scenario, the channel correlation matrices and the DL model are derived from channel samples generated at the same speed as the test condition. For the mismatched scenario, the DL model and correlation estimation are trained using samples obtained at a lower speed of $v = 72$ km/h, allowing us to evaluate the generalization capability of the proposed methods. 

Performance evaluations for the $2 \times 2$ MIMO-OFDM system are presented in \figref{fig_comp_mm}(a). Under the matched condition, both VMP-based approaches (VMP and VMP-L) and the proposed DL network perform comparably to the LMMSE-based refined channel estimation. In contrast, under the mismatched condition, the JCDD receiver using LMMSE-based refinement experiences notable performance degradation. This decline is primarily due to the use of mismatched channel statistics in the time-domain interpolation step defined in \eqref{equ_Almmse_time}, which results in inaccurate interference cancellation caused by unreliable feedback. The VMP-based methods overcome this issue by applying variational inference independently of time correlation and suppressing error propagation through the message passing mechanism. Meanwhile, the proposed DL network enhances generalization to varying channel conditions via convolutional feature fusion and mitigates feedback errors through the attention mechanism. As observed in \figref{fig_comp_mm}(a), these adaptive enhancements offer a performance gain exceeding 2 dB compared to ``$I=2$, LMMSE (Mismatch).''

\figref{fig_comp_mm}(b) presents performance comparisons for the $64 \times 4$ MIMO-OFDM system. In the matched scenario, both the VMP-L and DL designs offer performance similar to the LMMSE-based method. However, the original VMP algorithm becomes computationally intractable due to the high dimensionality associated with spatial-frequency domain concatenation. In the mismatched case, the performance improvement of VMP-L over LMMSE is marginal. This is mainly attributed to the limitations in the observation approximation as described in \eqref{equ_yhat_vmpl}. By contrast, the proposed DL network remains robust under statistical mismatches, delivering a performance gain of more than 2 dB over ``$I=2$, LMMSE (Mismatch)'' and approximately 1.5 dB over ``$I=2$, VMP-L (Mismatch).''

\vspace{-1em}
\subsection{Complexity Analysis}
We further examine the computational complexity of the refined channel estimation methods employed in the proposed JCDD receiver, as summarized in \tabref{tab_complexity}. For both the LMMSE and VMP-based estimators, most of the computational load arises from matrix inversion. The complexity of the LMMSE-based method is given by $\mathcal{O}( {{N_{\text{r}}}{N_{\text{t}}}( {{K^3}T + {T^3}K} )} )$, which accounts for both frequency- and time-domain interpolations. In contrast, the VMP algorithm involves high-dimensional matrix operations for joint estimation across all antennas and subcarriers, resulting in significantly higher complexity of $\mathcal{O}( {T{{( {{N_{\text{r}}}{N_{\text{t}}}K} )}^3}} )$, as shown in \eqref{equ_hhat_vmp}. To mitigate this, the VMP-L method applies a separate variational inference scheme for each antenna pair. This reduces the computational burden to $\mathcal{O}\left( {T{N_{\text{t}}}\left( {{N_{\text{r}}} + 1} \right){K^3}} \right)$ and makes the method more scalable for large MIMO configurations.

\begin{figure*}[t]
  \centering
  \includegraphics[width=0.74\textwidth]{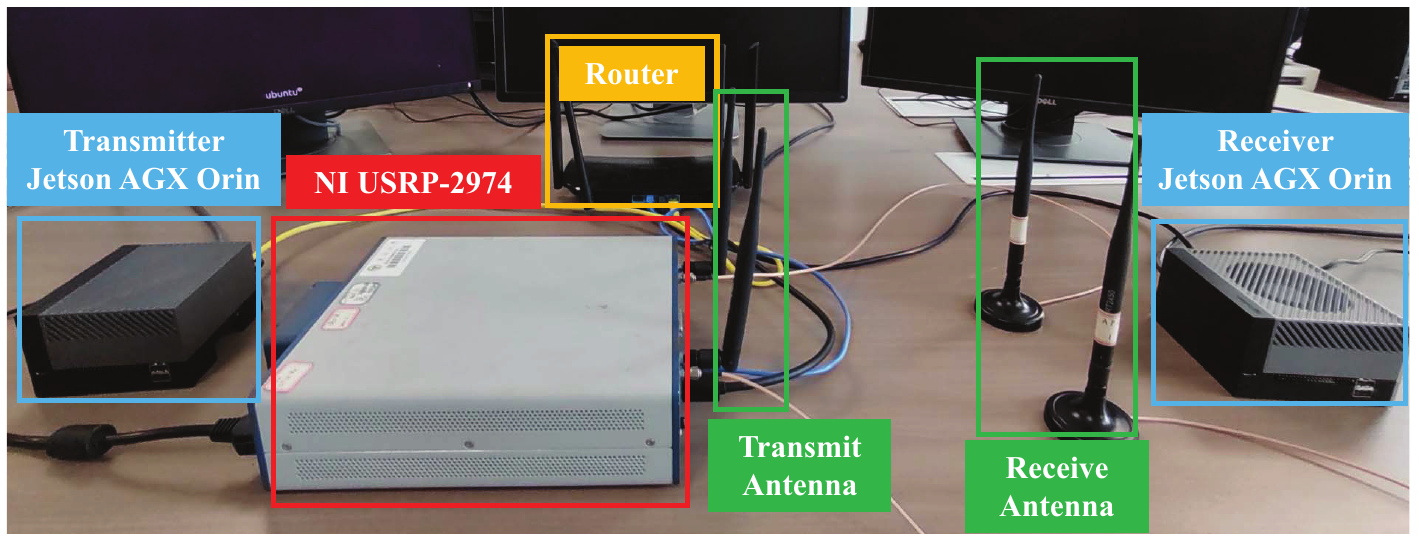}
  \caption{Hardware components of the prototype platform.}
  \label{fig_ota_system}
  \vspace{-1em}
\end{figure*}

The computational complexity of the proposed DL-based estimator is also provided in \tabref{tab_complexity}. The CNN-based architecture exhibits linear complexity growth with respect to the number of transmitted resource blocks. This contrasts with the cubic scaling observed in the LMMSE and VMP-based methods. With a parameter count of $6.87 \times 10^{4}$, the network remains lightweight and allows for straightforward extension to various MIMO configurations without modifying the model structure. Processing latency is evaluated for both $2\times2$ and $64\times4$ MIMO settings. The DL-based method achieves latency comparable to that of the VMP-L approach. The full VMP method introduces substantial delays due to its high-dimensional inference, and its latency in the $64\times4$ setting is omitted due to computational intractability. These findings are consistent with the earlier complexity analysis. In summary, the DL-based refined estimation used in the JCDD receiver with $I = 2$ iterations provides a favorable balance among adaptability, latency, and complexity, making it a promising solution for practical high-mobility deployment. 

\section{OTA Test}
\label{sec_ota}
To demonstrate the practical viability of the proposed SIP receivers, we conduct OTA  transmission experiments using a $2 \times 2$ MIMO-OFDM prototype platform. This section begins by presenting the system-level architecture. Next, the implementation details are described, with a focus on software design and the signal processing flow. Finally, we report experimental findings based on the OTA measurements.

\vspace{-1em}
\subsection{System Framework}
\figref{fig_ota_system} illustrates the system-level architecture of the prototyping platform. The platform consists of a high-performance computing unit for baseband processing and software-defined radios (SDRs) for RF transmission and reception. Baseband operations are executed on the NVIDIA Jetson AGX Orin (referred to as \textbf{Jetson}), which features a 2048-core Ampere GPU operating at up to 1.3 GHz and supports a peak computational throughput of 275 trillion operations per second (TOPS). The RF front-end uses the USRP-2974 from National Instruments (NI) (referred to as \textbf{USRP}), which integrates two RF channels and supports $2 \times 2$ MIMO transmission using a LabVIEW-based reconfigurable I/O architecture. The transmit and receive antenna arrays are placed 0.4 meters apart. 

\begin{figure}[t]
  \centering
  \includegraphics[width=0.5\textwidth]{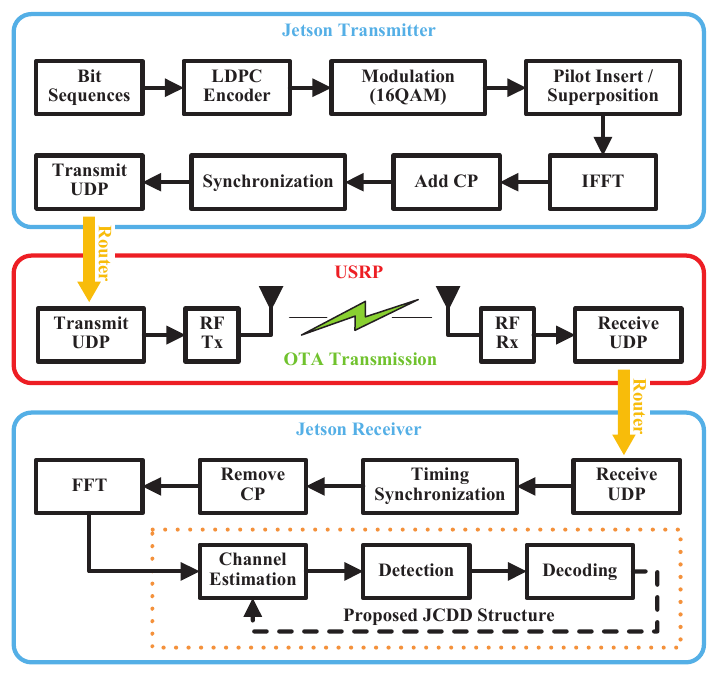}
  \caption{Diagram of the system flow chart.}
  \label{fig_ota_system_flow}
\end{figure}

The Jetson computing units and the USRP are connected to a local Ethernet network via a router, enabling data exchange using the User Datagram Protocol (UDP). As shown in \figref{fig_ota_system_flow}, the \textbf{Jetson transmitter} generates time-domain OFDM waveforms, encapsulates them into UDP packets, and continuously transmits these packets to the \textbf{router}. The \textbf{USRP} listens for incoming UDP packets, extracts the baseband signals, performs I/Q modulation, and upconverts the signals to a 2 GHz carrier frequency for OTA transmission using its antenna array. Simultaneously, the \textbf{USRP} executes a reception loop that amplifies the received RF signals, downconverts them to baseband, and digitizes the resulting waveforms. These baseband samples are then encapsulated into UDP packets and sent through the \textbf{router} to the \textbf{Jetson receiver}, which performs UDP unpacking, baseband signal processing, and demodulation to recover the transmitted information bits.

\begin{figure*}[!ht]
  \centering
  \includegraphics[width=0.7\textwidth]{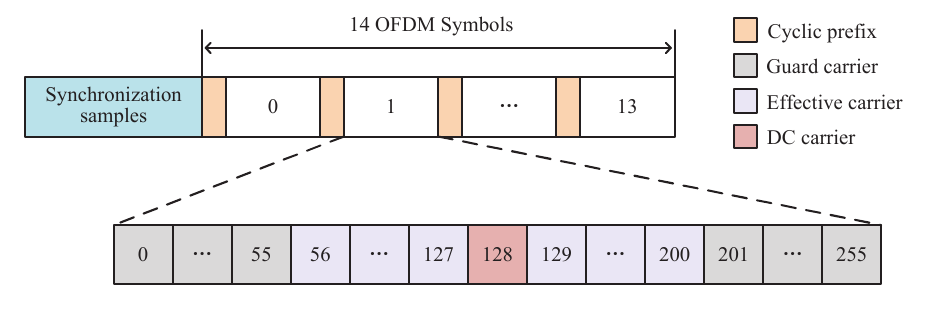}
  \caption{Frame structure used in the prototype system.}
  \label{fig_ota_frame}
  \vspace{-1em}
\end{figure*}

\subsection{Software Design}
The software implementation of the prototype system is realized using two primary programming environments: Python for the Jetson processors and LabVIEW for the USRP hardware, enabling physical-layer signal processing. The frame structure used in OTA transmission is shown in \figref{fig_ota_frame}. It comprises a synchronization sequence followed by 14 OFDM symbols. The synchronization sequence consists of 2048 samples and is used to facilitate timing synchronization. Upon receiving the signal, the receiver performs correlation with the known synchronization pattern to identify the start index of the data payload within each frame.

\begin{table}[!ht]
  \caption{Parameters for the Prototype System}
  \label{tab_ota_parameter}
  \centering
  \begin{tabular}{ll}
    \toprule
    Parameter & Value \\
    \midrule
    MIMO & $2\times2$ \\
    Carrier Frequency & 2 GHz \\
    Sampling Frequency & 7.68 MHz \\
    System Bandwidth & 2.37 MHz \\
    Subcarrier spacing & 30 kHz \\
    FFT size & 256 \\
    Symbols per Frame & 14 \\
    Guard Carriers & (55,56) \\
    DC Carrier & True \\
    CP Length & 64 \\
    Synchronization Samples & 2048 \\
    OFDM Symbol Duration & 41.67 $\mu$s \\
    Frame Duration & 0.58 ms \\
    Number of Frames & 10000 \\
    \bottomrule
  \end{tabular}
\end{table}

Each OFDM symbol consists of 256 subcarriers with an appended 64-point cyclic prefix (CP) to mitigate inter-symbol interference. Within each OFDM symbol, 144 subcarriers are allocated for pilot and data transmission, 111 subcarriers are reserved as guard bands, and one subcarrier is dedicated to DC offset. We consider a static scenario with $v = 0$ km/h, and the OP transmission adopts the ``1P'' pilot pattern, as depicted in \figref{fig_op_pilot}(a). The detailed system parameters of the prototyping platform are listed in \tabref{tab_ota_parameter}.

At the transmitter, the frequency-domain OFDM symbol is converted into the time domain using an inverse fast Fourier transform (IFFT), as presented in \figref{fig_ota_system_flow}. At the receiver, after timing synchronization, OFDM demodulation recovers the frequency-domain signals. These are then processed by subsequent modules including channel estimation, signal detection, and decoding. The modulation and coding schemes, as well as the power allocation ratio, are kept identical to those used in the simulation configuration in \secref{sec_simulation}. The proposed iterative SIP receiver employs the VMP, VMP-L, and DL-based refined estimation techniques. To assess performance, the demodulated bit sequences are compared with the original transmitted bits to compute BLER and throughput metrics, thereby validating the proposed framework. 

\vspace{-1em}
\subsection{Experimental Results}

\begin{figure}[t]
  \centering
  \begin{minipage}{3.5in}
    \centerline{\includegraphics[width=3.5in]{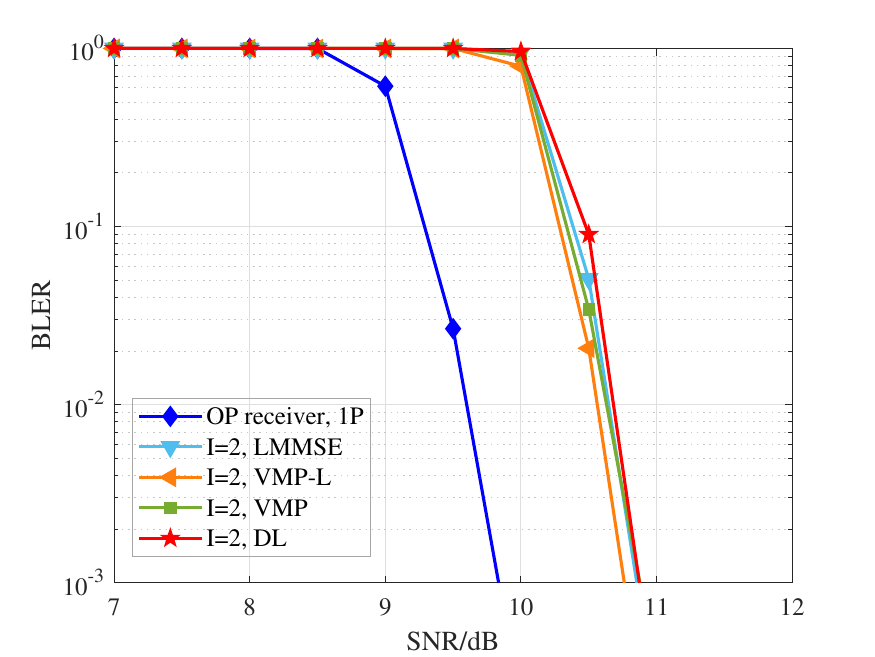}}
    \centerline{\small{(a) BLER}}
  \end{minipage}
  \hfill
  \begin{minipage}{3.5in}
    \centerline{\includegraphics[width=3.5in]{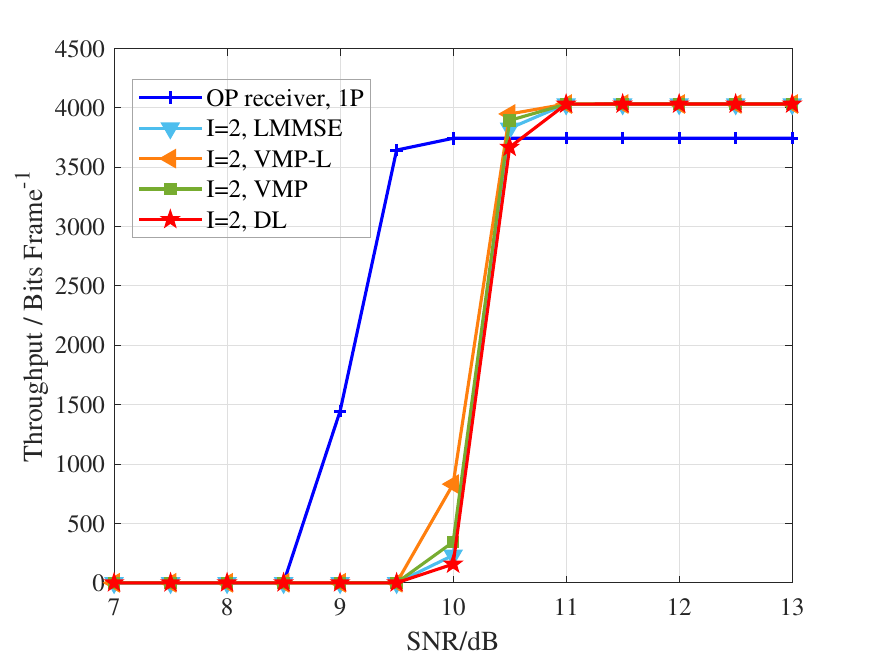}}
    \centerline{\small{(b) Throughput}}
  \end{minipage}
  \caption{Performance comparisons between the OP receiver and our proposed SIP receivers under OTA transmissions.}
  \label{fig_comp_ota}
\end{figure}

Due to the presence of a line-of-sight (LOS) path, the OTA channel exhibits statistical characteristics that differ from those of the CDL-C model used in simulation studies. Therefore, the correlation matrices used for channel estimation are computed based on second-order statistical averages over $10^4$ LS-based channel estimates collected from OTA measurements using the OP transmission scheme. For deployment of the proposed DL-based estimation framework, the model is trained on $5 \times 10^5$ OTA samples collected under the SIP transmission scheme. This ensures adaptation to realistic channel environments. 

The performance evaluation results are illustrated in \figref{fig_comp_ota}. According to \figref{fig_comp_ota}(a), under matched channel statistics, the adaptive enhancement strategies achieve performance comparable to that of the LMMSE-based method within the proposed iterative receiver. In this case, the OP receiver demonstrates a BLER advantage of approximately 1 dB compared with the iterative SIP receiver. However, despite this slight gain in estimation accuracy, our proposed scheme achieves a 7.7\% improvement in throughput, as shown in \figref{fig_comp_ota}(b). These findings are consistent with the simulation results under low-mobility conditions, thereby confirming the practical viability of our developed iterative receiver.

\section{Conclusion}
\label{sec_conclusion}
This paper proposed an iterative IC-based JCDD receiver for SIP transmission in MIMO-OFDM systems. To address the challenges posed by pilot contamination and data interference, we introduced adaptive refined channel estimation strategies, including a VMP method, its low-complexity variant (VMP-L), and a DL-based approach. The VMP-based estimators eliminate the need for time-domain correlation statistics, while the DL-based design leverages convolutional and attention-based mechanisms to capture channel features and mitigate statistical mismatch.
Extensive simulations under various antenna configurations and mobility scenarios demonstrated that the proposed receiver achieves significant throughput gains over traditional OP-based baselines, while maintaining comparable or improved BLER performance. In particular, the proposed DL-based estimator showed strong generalization capability with a favorable trade-off between accuracy and complexity. Furthermore, OTA experiments confirmed the practical effectiveness of the proposed design in real-world wireless channels.

\ifCLASSOPTIONcaptionsoff
  \newpage
\fi


\end{document}